\documentclass[11pt,twocolumn]{emulateapj}

\tightenlines
\usepackage{epstopdf}
\usepackage{ulem} 
\usepackage{color} 
\usepackage{lineno}

\def\lsim{\;\raise0.3ex\hbox{$<$\kern-0.75em\raise-1.1ex\hbox{$\sim$}}\;}
\def\gsim{\;\raise0.3ex\hbox{$>$\kern-0.75em\raise-1.1ex\hbox{$\sim$}}\;}

\definecolor{purple}{RGB}{200,100,255} 
\definecolor{orange}{RGB}{255,69,0}

\newcommand{\Bhom}{B_\mathrm{hom}}
\newcommand{\Btur}{B_\mathrm{tur}}
\newcommand{\BturSq}{\langle \Btur^2 \rangle}
\newcommand{\Lmin}{L_\mathrm{min}}
\newcommand{\Lmax}{L_\mathrm{max}}
\newcommand{\Rsnr}{R_\mathrm{SNR}}

\newcommand{\Brms}{\langle B^2 \rangle}

\newcommand{\gamray}{$\gamma$-ray}
\newcommand{\gamrays}{$\gamma$-rays}

\newcommand{\syn}{synchrotron}

\newcommand{\RT}{Raleigh--Taylor}
\newcommand{\KH}{Kelvin--Helmholtz}

\newcommand{\rel}{relativistic}

\newcommand{\alf}{Alfv\'en}

\newcommand{\Alfic}{Alfv\'enic}

\newcommand{\NL}{nonlinear}

\newcommand{\chandra}{{\sl Chandra}}
\newcommand{\ixpe}{{\sl IXPE}}

\renewcommand{\baselinestretch}{1}
\def\aap{Astronomy and Astrophysics}

\newcount\listnorom
\newcommand\listromanDE{\global\advance \listnorom by 1
{\lowercase\expandafter{(\romannumeral\listnorom)}\ }}

\newcount\listnumber
\newcommand\listDE{\global\advance \listnumber by 1
{\lowercase\expandafter{(\number\listnumber)}\ }}

\newcount\Inum
\newcount\IInum
\newcount\IIInum
\newcount\IVnum
\def\I{\global\multiply\IInum by 0 \global\multiply\IIInum by 0
            \global\multiply\IVnum by 0 \global\advance \Inum by 1
            {\the\Inum. }}
\def\II{\global\multiply\IIInum by 0\global\multiply\IVnum by 0
       \global\advance \IInum by 1 {\the\Inum.\the\IInum. }}
\def\III{\global\multiply\IVnum by 0\global\advance \IIInum by 1
            {\the\Inum.\the\IInum.\the\IIInum. }}
\def\IV{\global\advance \IVnum by 1
            {\the\IVnum. }}
\shorttitle{Polarized X-rays in SNRs}
\shortauthors{Bykov, Uvarov, Slane \& Ellison}

\begin{document}

\title{Uncovering magnetic turbulence in young supernova remnants with polarized X-ray imaging}

\vskip24pt

\author{
Andrei M. Bykov,\altaffilmark{1},
Yury A. Uvarov,\altaffilmark{1},
Patrick Slane,\altaffilmark{2}
and Donald C. Ellison\altaffilmark{3} }

\altaffiltext{1}{Ioffe Institute for Physics and Technology, 194021
St. Petersburg, Russia; byk@astro.ioffe.ru}


\altaffiltext{2}{Harvard-Smithsonian Center for Astrophysics,
Cambridge MA 02138, U.S.A; slane@cfa.harvard.edu}

\altaffiltext{3}{Physics Department, North Carolina State
University, Box 8202, Raleigh, NC 27695, U.S.A.;
ellison@ncsu.edu}


\begin{abstract}
Observations of young supernova remnants (SNRs) in X-rays and
\gamrays\ have provided conclusive evidence for particle acceleration
to at least TeV energies.
Analysis of high spatial resolution X-ray maps of young SNRs has
indicated that the particle acceleration process is accompanied by
strong non-adiabatic amplification of magnetic fields.
If Fermi acceleration is the mechanism producing the energetic cosmic
rays (CRs), the amplified magnetic field must be turbulent and 
CR-driven instabilities are among the most probable mechanisms for
converting the shock ram pressure into the magnetic turbulence.
The development and evolution of strong magnetic turbulence in the
collisionless plasmas forming SNR shells are complicated phenomena which
include the amplification of magnetic modes, anisotropic mode transformations at
shocks, as well as the nonlinear physics of turbulent cascades.
Polarized X-ray synchrotron radiation from  ultra-relativistic electrons
accelerated in the SNR shock is produced in a thin layer immediately
behind the shock and is not subject to the Faraday depolarization effect.
These factors open possibilities to study some properties of magnetic
turbulence and here we present polarized X-ray synchrotron maps of SNR
shells assuming different models of magnetic turbulence cascades.
It is shown that different models of the anisotropic turbulence 
can be distinguished by measuring the predominant polarization angle
direction. We discuss the detection of these features in Tycho's SNR
with the coming generation of X-ray polarimeters such as \ixpe.
\end{abstract}

\section{Introduction}
The synchrotron interpretation of 
non-thermal radio emission from ``radio stars"   
by \citet[][]{1950PhRv...78..616A}, 
while not directly mentioning supernovae (SNe), gave observational support to the hypothesis by \citet{BZ34} that SNe are 
sources of Galactic cosmic rays 
\citep[see e.g.][]{GS64,Axford81,2014IJMPD..2330013A}. 
Collisionless shocks produced by supersonic SN ejecta expanding into the interstellar medium, simultaneously accelerate charged particles by the diffusive shock acceleration (DSA) mechanism while amplifying the local turbulent magnetic field 
\citep[for reviews, see][]{Blandford_Eichler_1987,SchureEtal2012,Marcowith2016}.
As a result, quasi-power law particle spectra are formed where a significant fraction of the total SN ejecta kinetic energy can be deposited in \rel\ cosmic rays (CRs). 

Electrons can
be accelerated by fast shocks up to TeV energies and efficiently emit synchrotron radiation up to X-ray 
energies in the amplified magnetic fields \citep[see e.g.][]{Reynolds_Chevalier_1981}.
This model is supported by observations which confirm the existence
of a power law component in nonthermal SNR radiation in a broad
spectral range from radio to X-rays \citep[see e.g.][]{helder12}.  

Direct observations of polarized radio emission from SNRs proved the synchrotron nature of the radiation and provided a way to study the magnetic field structure \citep[see][for reviews]{1977ARA&A..15..175C,1984ARA&A..22...75R,2012SSRv..166..231R,2015A&ARv..23....3D}.
Radio mapping of the young SNRs Cas A \citep{1975Natur.257..463B} and Tycho
\citep[][]{1973A&A....25..351S,1975A&A....39...33D,1991AJ....101.2151D,1997ApJ...491..816R} resolved
well-defined edges of the radio shells associated with the forward shock waves.
This magnetic field structure revealed from radio observations in young SNRs is often characterized by predominantly radial fields, while older SNRs, like the Cygnus Loop, IC 443 and others, show a tangential magnetic field structure \citep[e.g.,][]{2004mim..proc..141F}.  

The nature of the apparent predominantly radial field is still under discussion. 
Recently, \citet[][]{2017ApJ...849L..22W} considered 
selection effects due to the distribution of CR electrons accelerated by quasi-parallel shocks as an explanation 
of the radial magnetic field pattern derived from radio polarization observations.
The authors demonstrated that radio-emitting electrons accelerated at quasi-parallel shocks can have spatial distributions that result in an apparent radial magnetic field  in radio synchrotron maps even when the field is in fact disordered.
\citet{2016MNRAS.459..178B} and \citet{2017MNRAS.470.1156P} modeled radio images of SNRs accounting for turbulent 
magnetic fields. 

Since GeV electrons, regardless of origin,  
have long lifetimes in the Galaxy, their synchrotron radio emission can be used to study large scale magnetic turbulence 
\citep[][]{2019ApJ...877..108L,2020ApJ...890...70W}.
The ubiquitous nature of radio emission in SNRs provides direct constraints on the injection and shock acceleration of GeV electrons, as well as propagation and polarization
in turbulent magnetic fields 
\citep[e.g.,][]{Lazendic2004,Reynolds08}. Here, however, we limit discussion to polarized X-ray \syn\ radiation where 
the rapid energy loss of TeV electrons means they are restricted to a thin shell behind the 
remnant forward shock.  In this case,  the Faraday depolarization effects are negligible in X-rays contrary to the radio band observations.

This provides the opportunity to reach a degree of polarization detectable
with the new generation of X-ray polarimeters such as 
the Imaging X-ray Polarimetry Explorer (\ixpe) 
\citep[]{Weisskopf_2016}.  

The TeV electrons producing X-ray \syn\ emission in SNRs are likely accelerated by the 
DSA mechanism 
\citep[e.g.,][]{Blandford_Eichler_1987,ESPB2012}. 
This mechanism requires  magnetic turbulence to function and, while pre-existing background turbulence is always present in astrophysical plasmas, locally generated broadband turbulence is 
necessary to produce GeV-TeV CRs. 
This turbulence is most likely produced in the shock precursor by instabilities generated by
backstreaming CRs.
Models show that these instabilities can transfer a few percent of the shock ram pressure to the magnetic field fluctuations \citep[see e.g.][]{Bell2004,AB2006,SchureEtal2012,Bykov3inst2014}.
The effect of turbulent magnetic fields on X-ray synchrotron images of young SNRs was discussed by \citet{Bykov2008,Bykov2009}.

High resolution \chandra\ X-ray \syn\ maps of young SNRs  revealed a 
number of structures of 
different sizes and morphologies \citep{vink12}.
These included narrow filaments, stripes, and clumps 
\citep[see e.g.][]{Vink_Laming_2003,Bamba_2005,PF2009,EriksenEtal2011}.
It was also discovered that small scale structures were
time-variable on a scale of a few years 
in the case of SNR RX J1713.72 \citep{Uchiyama2007} 
Cas A \citep{UA2008}, and Tycho \citep{2020arXiv200310035O}.
It was argued that the filamentary structures might result from fast energy losses of the X-ray emitting TeV electrons downstream from the shock if the downstream field was  amplified well above the adiabatic compression magnitude.
If so, the filaments are due to the geometric projection of the thin regions accelerating the TeV electrons. An alternative explanation of the filaments is due to a rapid decay of amplified magnetic fields in the vicinity of the shock wave \citep{Pohl2005}.

The time-variable clumpy structure detected by \chandra\ 
can also be naturally 
explained as a result of turbulent
magnetic fields in the vicinity of the shock front  without postulating an exceptionally strong uniform background field or field decay
\citep[][]{Bykov2008}. 
In  this  model, the intensity and polarization of the structures are sensitive to the
magnetic turbulence spectrum even in the case of isotropic turbulence \citep{Bykov2009}.
The observational diagnostics of the isotropic turbulence in SNRs will require 
arcsecond resolution polarimetry which could be available with the next generation of X-ray 
polarimeters.

However, as we show below,  anisotropic turbulence in  
young SNRs can be tested with upcoming polarimeters like \ixpe. 
Not only is broadband self-generated turbulence necessary to produce GeV-TeV CR electrons and ions, there is ample evidence, particularly from sharp X-ray \syn\ edges,  that this turbulence can be amplified by orders of magnitude above 
background levels \citep[e.g.,][]{Parizot2006}.
As this turbulence is compressed upon entering the downstream region, 
where most of the X-ray \syn\ radiation is produced,
it should become highly 
anisotropic.\footnote{We note that ambient magnetic fields can be amplified by non-CR processes such as \RT\ and \KH\ instabilities and this may  result in a net radial polarization behind the forward shock \citep[e.g.,][]{JunNorman1996}.}

In addition, the  turbulence anisotropy 
may evolve due to MHD plasma
evolution \citep{1983JPlPh..29..525S,2002ApJ...564..291C,2008PhRvE..78f6301B,Reville_2008}, such as 
anisotropic wave cascading \citep{Goldr1994,Goldr1997,Lith2003}.
In this paper we consider both (i) anisotropic turbulence due to shock compression of isotropic precursor turbulence\footnote{In oblique shocks, where the large-scale magnetic field direction is oblique to the shock normal, the magnetic field component tangent to the shock front is increased upon shock compression.} 
and (ii) the anisotropic turbulence emerging
in a cascading process. 
These two scenarios produce turbulence with different characteristics that can be distinguished in SNR X-ray \syn\ maps.

\section{Modeling of SNR synchrotron X-ray images}
%
\subsection{Turbulence cascade}

The time evolution of a turbulent magnetized plasma is a
nonlinear process. However, in some conditions, the complicated physics of this process can be described as an interaction of linear MHD waves.
In \citet{Goldr1994} it was shown for weak
turbulence (i.e., when the small scale turbulent fields are weaker than 
the large scale field) in incompressible magnetized fluids that the  
3-\alf-wave interaction
is absent and a 4-wave interaction leads to the 
generation of waves with
higher perpendicular wave number $k_{\perp}$.  
A special turbulent spectrum is formed with energy density 
$Wd^{3}k=C\left(k_{\parallel}\right)k_{\perp}^{-10/3}dk_{\parallel}k_{\perp}dk_{\perp}$. 
While the spectrum has a specific dependence on $k_{\perp}$, the dependence on 
$k_{\parallel}$ is determined by the initial turbulence.

If energy is injected in turbulence on large scales then only small
$k_{\parallel}$ are present in the turbulent spectrum and the assumption 
$C\left(k_{\parallel}\right)\approx {\rm const}$ can be used.
%
%
If the initial turbulence is isotropic 
(i.e., $\langle\delta B_{x}^{2}\rangle=
\langle\delta B_{y}^{2}\rangle=\langle\delta B_{z}^{2}\rangle=\langle\delta B^{2}\rangle/3$)
the evolved turbulence will have 
$\langle\delta B_{\parallel}^{2}\rangle>\langle\delta B_{\perp}^{2}\rangle/2$, where $\langle\delta B_{\parallel}^{2}\rangle$ is the mean square of the turbulent
magnetic field component directed along the large scale magnetic field and $\delta B_\perp$ is the transverse component of the turbulent field.
%
This effect is a result of the evolution of \Alfic\ 
waves polarized
so their magnetic field lies in the same plane as a wave vector and
a large scale field. The existence of such polarization of \alf\
waves is a consequence of incompressibility.

\begin{figure}
\vskip12pt
\epsscale{1.0}
\plotone{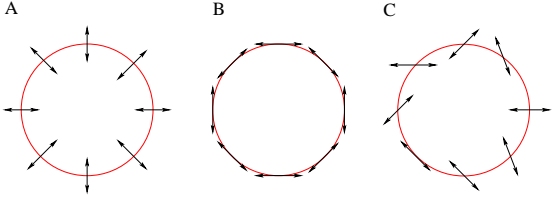}  
\caption{Sketch of SNR forward shock showing polarization directions immediately downstream. 
Example (A) shows the case of anisotropic turbulence
produced by shock compression of upstream isotropic turbulence. 
Example (B) shows the case of
turbulence produced by the anisotropic cascade of an almost radial magnetic field near the shock.
Example C) shows the domain structure formed from long-scale stochastic fluctuations
of CR-driven magnetic turbulence  produced by hard spectrum CRs in \NL\ DSA (see text).
\label{fig:sketch}}
\end{figure}

In this work we consider three possible scenarios for polarization directions in supernova shells as sketched in Fig.~\ref{fig:sketch}.
In (A) we show the case where there is strong, isotropic, short-scale turbulence in the shock precursor. Upon compression as the turbulence is swept across the shock,  the magnetic field will become 
predominantly tangential to the shock front and the polarization will be predominantly radial. 
One should note that the \NL\ interactions of the fluctuations downstream from the shock  
may isotropise the tangential anisotropy induced by the transition through an oblique shock and reduce the degree of X-ray polarization. 
However, due to fast losses of TeV electrons the synchrotron X-ray radiation is formed in a thin ridge just behind the shock where the turbulence isotropization effect may be incomplete and significant X-ray polarization can be detected.\footnote{We illustrated both the fully anisotropic and fully isotropic cases in  Fig.~\ref{fig:1.2_isotr_and_anisotr}.}

In the case of anisotropic cascades (B), the predominant direction of the turbulent field is the same as the large-scale field. Since  large-scale fields in young SNRs tend to be radial at the forward shock, 
\syn\ radiation will have  a polarization direction predominantly tangent to the shock.

Our third case (C) assumes the CRs accelerated in the forward shock have a hard enough  spectrum so strong, long-wavelength fluctuations on spatial scales up to 10~arcsec are produced in Tycho's SNR.  It should be noted here that 
even in the cosmic ray driven turbulence the long-wavelength fluctuations may have scales larger than the gyroradii of the highest energy cosmic rays if non-resonant instablities like the firehose \citep[see e.g.][]{SchureEtal2012,bbmo13} or mirror instabilities \citep[][]{Bykov_SuperD2017} are in operation.  
In this case, the magnetic field near the shock will form domains 
of 
this size
with almost random field and polarization directions. 
The life-time of such long-scale fluctuations is longer than 
a year.

\subsection{Numerical simulation of an anisotropic stochastic magnetic field}

In a cascading process the direction of the large scale magnetic field
defines the symmetry axis of the  problem. The general
stochastic property of axially symmetric magnetic turbulence
is 
\begin{equation}
\langle B_{\perp}^{2}\rangle=q\langle B_{\parallel}^{2}\rangle=\langle B^{2}\rangle q/\left(q+1\right)
\ .
\end{equation}
%
In the case of isotropic turbulence $\langle B_{x}^{2}\rangle=\langle B_{y}^{2}\rangle=\langle B_{z}^{2}\rangle=\langle B^{2}\rangle/3$
and $q=2$.
For turbulence formed downstream from the shock by shock compression of initially isotropic turbulence $q>2$, while $q<2$ for  turbulence produced in the cascading process
described above.

For our simulation of anisotropic turbulence we use 
a  method based on the work of 
\citet{Giakalone_Jokipii_1999}. We consider the summation
of a large number of harmonics with random wave vectors and phases:
\begin{equation}
{\bf B}({\bf r},t)=\sum_{n=1}^{N_{m}}\sum_{\alpha=1}^{2}{\bf A}^{(\alpha)}(k_{n})\cos({\bf k}_{n}\cdot{\bf r}-\omega_{n}({\bf k}_{n})\cdot t+\phi_{n}^{(\alpha)})
\end{equation}
We have two orthogonal polarizations ${\bf A}^{(\alpha)}(k_{n})$
($\alpha=1,2$) of magnetic field, both chosen orthogonal to the wave
vector ${\bf A}^{(\alpha)}(k_{n})\perp{\bf k}_{n}$ in order to satisfy
$\nabla\cdot{\bf B}=0$.
%
For the 1st polarization we choose
the direction of the magnetic field to lie in a plane formed
by the symmetry
axis and the wave vector. 
The magnetic field direction for the 2nd polarization is orthogonal to this plane. We introduce here the coordinate system which is consisting
of the axis $||$ directed along the symmetry axis and the perpendicular
plane. 
The 2nd polarization gives contribution only to the value of
$B_{\perp}$, while the 1st polarization contributes  both to
$B_{\perp}$ and $B_{\parallel}$. The amplitudes for polarizations
1,2 are 
\begin{equation}
A_{1}^{2}({\bf k})=\left(1-a\right)A^{2}({\bf k})
\ ,
\end{equation}
\begin{equation}
A_{2}^{2}({\bf k})=\left(1+a\right)A^{2}({\bf k})
\ ,
\end{equation}
where
$a\geq-1$ and $q=\left(2+a\right)/\left(1-a\right)$. 
In the case of isotropic turbulence, the parameter $a=0$ ($q=2$) and the 
spectral energy
density of the magnetic turbulence is
\begin{equation}
W d^{3}k = C k^{-\delta} dk =
4 \pi A^2(k) \cdot k^2 dk
\ .
\end{equation}
In the case of anisotropic turbulence produced by the shock compression of 
isotropic turbulence, we model the spectral energy density with the same expression 
with $a>0$ ($q>2$). For the anisotropic cascades considered by \citet{Goldr1994} 
the spectral energy density is
\begin{equation}
Wd^{3}k=C\left(k_{\parallel}\right)k_{\perp}^{-10/3}dk_{\parallel}k_{\perp}dk_{\perp}=2\pi A^{2}(k)dk_{z}k_{\perp}dk_{\perp}.
\end{equation}

\subsection{Synchrotron radiation properties}

A 2D SNR map  is obtained by integrating a 3D distribution of \syn\ emissivity along each line-of-sight.
We consider magnetic
field fluctuations with  characteristic  sizes greater than the formation
length $l_{f}\approx mc^{2}/eH\approx1.7\times10^{9}B_{\mu G}^{-1}$
of the synchrotron radiation. This allows use of the  standard formula for synchrotron
radiation in a homogeneous magnetic field as given by \citet{Ginzburg1965}. 

The Stokes parameters,
{$\tilde{I},\, \tilde{Q},\, \tilde{U},\, {\rm and}\,  \tilde{V}$},
that fully describe radiation polarization properties
can be written as
\begin{equation}
\widetilde{I}({\bf r},t,\nu)= S_c 
\, B_{\perp}\left({\bf r}\right)\int\limits _{\nu/\nu_{c}}^{\infty}K_{5/3}(\eta)d\eta
\ ,
\end{equation}
\begin{equation}
\widetilde{Q}({\bf r},t,\nu)= S_c
\, B_{\perp}\left({\bf r}\right)K_{2/3}\left(\frac{\nu}{\nu_{c}}\right)\cos{(2\chi)}
\ ,
\end{equation}
\begin{equation}
\widetilde{U}({\bf r},t,\nu)= S_c 
\, B_{\perp}\left({\bf r}\right)K_{2/3}\left(\frac{\nu}{\nu_{c}}\right)\sin{(2\chi)}
\ ,
\end{equation}
\\
where $S_c=[\sqrt{3} e^3/(mc^2 r^2)] (\nu/\nu_c)$ and 
$\tilde{V} =0$ for the isotropic particle distributions we consider here.
These parameters
are additive for incoherent photons so maps of Stokes parameters can
be obtained by integration over the line-of-sight taking into account
the distribution of emitting 
particles $f_{e}\left(E,{\bf r}\right)$ and the expression 
$dV = r^2 dr d\Omega$:
\begin{eqnarray}
& & I(\nu) = \int I(\nu,{\bf r})dr= \\
& &
\frac{\sqrt{3}e^{3}}{mc^{2}}\int drdE\frac{\nu}{\nu_{c}}\, B_{\perp}\left({\bf r}\right)f_{e}\left(E,{\bf r}\right)\int\limits _{\nu/\nu_{c}}^{\infty}K_{5/3}(\eta)d\eta
\nonumber
\end{eqnarray}
\begin{eqnarray}
& & Q(\nu) = \int Q(\nu,{\bf r})dr= \\
& &
\frac{\sqrt{3}e^{3}}{mc^{2}}\int drdE\frac{\nu}{\nu_{c}}\, \cos{(2\chi)}
B_{\perp}\left({\bf r}\right)f_{e}\left(E,{\bf r}\right)K_{2/3}\left(\frac{\nu}{\nu_{c}}\right)
\nonumber
\end{eqnarray}
\begin{eqnarray}
& & U(\nu) = \int U(\nu,{\bf r})dr= \\
&  &
\frac{\sqrt{3}e^{3}}{mc^{2}}\int drdE\frac{\nu}{\nu_{c}}\, \sin{(2\chi)}
B_{\perp}\left({\bf r}\right)f_{e}\left(E,{\bf r}\right)K_{2/3}\left(\frac{\nu}{\nu_{c}}\right)
\nonumber
\end{eqnarray}
where
\begin{equation}
\nu_{c}=\frac{3eB_{\perp}}{4\pi mc}\gamma^{2}
\end{equation}
and
\begin{equation}
\int dEd\Omega_{E}\cdot f_{e}\left(E,{\bf r}\right)=4\pi\int dE\cdot f_{e}\left(E,{\bf r}\right)=n({\bf r})
\ .
\end{equation}
In these equations, $I,\, Q,\, {\rm and}\, U(\nu)$ are normalized so the radiation flux near
Earth is given by 
$dF(\nu)=I(\nu)d\Omega=(dS/r^2)I(\nu)$
and so on, $f_{e}$ is an isotropic electron distribution function,
${\bf B_{\perp}}$ is a magnetic field projection to a plane transverse
to the line-of-sight, and $\chi$ is the angle between the fixed
direction in this plane and the main axis of the
polarization ellipse.
The parameters $\nu_{c}$ and $\chi$ are functions of ${\bf r}$. 

Using the Stokes parameters, we obtain the degree of polarization as
\begin{equation}
\Pi(x,y)=\frac{\sqrt{U^{2}(x,y)+Q^{2}(x,y)}}{I(x,y)}
\ ,
\end{equation}
where the coordinate system is shown in Fig.~\ref{fig:geometry}.
We note that for large sources time delay is important and we take this 
into account with
\begin{equation}
\left(\begin{array}{c}
I\\
Q\\
U
\end{array}\right)(x,y)=\int\left(\begin{array}{c}
I\\
Q\\
U
\end{array}\right) \left( x,y,z,t - \frac{z}{c} \right) dz
\ .
\end{equation}

The main axis of the polarization ellipse of emitting synchrotron
radiation is locally perpendicular  to the  magnetic
field projection $B_{\perp}\left({\bf r}\right)$. If the rotation
angle of polarization ellipse, $\chi$, is measured from the $-Ox$
axis then $\cos{\chi}=B_{\perp y}/B_{\perp}$ and $\sin{\chi} =B_{\perp x}/B_{\perp}$,
so 
\begin{equation}
\widetilde{Q}({\bf r},t,\nu) = S_c
B_{\perp}\left({\bf r}\right)K_{2/3}\left(\frac{\nu}{\nu_{c}}\right)\frac{B_{\perp y}^{2}-B_{\perp x}^{2}}{B_{\perp}^{2}}
\ ,
\end{equation}
\begin{equation}
\widetilde{U}({\bf r},t,\nu)= S_c
B_{\perp}\left({\bf r}\right)K_{2/3}\left(\frac{\nu}{\nu_{c}}\right)\frac{2B_{\perp y}\cdot B_{\perp x}}{B_{\perp}^{2}}
\ .
\end{equation}

It is easily seen that in the case of isotropic turbulence $\left\langle \widetilde{Q}({\bf r},t,\nu)\right\rangle =0$ and
$\left\langle \widetilde{U}({\bf r},t,\nu)\right\rangle =0$. 
In reality, 
the stochastic ensemble is not full so the average, while small, is not exactly zero. 
In the anisotropic case, if the axis of symmetry is perpendicular
to the line-of-site and coincides with axis Ox (or Oy) 
then the $x$ and $y$ field projections have different stochastic properties, i.e.,$\left\langle B_{\perp y}^{2}\right\rangle \neq\left\langle B_{\perp x}^{2}\right\rangle $
and $\left\langle \widetilde{Q}({\bf r},t,\nu)\right\rangle \neq0$,
while $\left\langle \widetilde{U}({\bf r},t,\nu)\right\rangle =0$.
If $B_{\perp}^{2}/B_{\parallel}^{2}$ is large then $\left\langle \widetilde{Q}({\bf r},t,\nu)\right\rangle $
can be close to the value for a homogeneous magnetic field
and the polarization can be close to the maximum theoretical limit.

\subsection{Electron distribution function}

Due to rapid radiation losses, high-energy, shock accelerated electrons
will be concentrated near the shock front.
The characteristic width of this region
is determined by the balance between the diffusive shock acceleration rate, diffusion out of the region, and the synchrotron loss rate. This width decreases rapidly
with increasing electron energy making a plane-shock approximation
reasonable. Following \citet{Bykov_2000} we write the plane-shock
diffusion-convection equation for DSA:
\begin{equation}
D\left(p\right)\frac{\partial^{2}f_e}{dx^{2}}-u\left(x\right)
\frac{\partial f_e}{\partial x}+\frac{p}{3}\frac{\partial u}{\partial x}\frac{\partial f_e}{\partial p}-\frac{1}{p^{2}}\frac{\partial}{\partial p}\left(p^{2}\frac{dp}{dt}f_e\right)=0\label{eq:distrFun}
\end{equation}
where
\begin{equation}
\frac{dp}{dt}=-\frac{32\pi}{9}\left(\frac{e^{2}}{m_{e}c^{2}}\right)^{2}\left(\frac{B^{2}}{8\pi}\right)\left(\frac{p}{m_{e}c}\right)^{2}
\end{equation}
describes synchrotron energy loses. For the plane-shock geometry,
$f_e(E,{\bf r}) = f_e[E,d({\bf r})]$,
$\int f_{e}\left(E,{\bf r}\right)dEd\Omega_{E}=4\pi\int f_{e}\left(E,{\bf r}\right)dE=n({\bf r})$,
and 
$d=|{\bf r}-{\bf R_{0}}|-\Rsnr$
is the distance to the shock front.
Here, ${\bf R_{0}}$ is a radius-vector
to the SNR center and $\Rsnr$ is SNR radius.

\begin{figure}
\vskip12pt
\epsscale{1.0}
\plotone{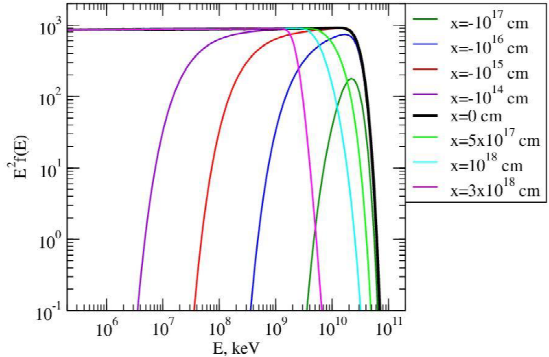}  
\caption{Electron distribution functions at different distances from
the shock front as indicated. Negative values correspond to the upstream region
while positive ones correspond to the downstream region.
\label{fig:model1}}
\end{figure}

For our simulations we calculated synchrotron
loses for a magnetic field $B = \sqrt{\Brms} =5.5\cdot10^{-5}$~G.
This value is justified if the characteristic size of the magnetic field fluctuations
is less than the average distance a particle moves during deceleration.
The forward shock velocity used was $2\cdot10^{8}$cm/s and we
assumed Bohm diffusion. The model distribution functions of electrons 
are shown in Fig.~\ref{fig:model1}. 
These electron distributions are generally consistent with standard
DSA models where ultra-relativistic electrons and protons have the same spectral shape up to the break in the electron spectra due to synchrotron losses.
While not shown, we find the  associated proton distributions in the
100 GeV-TeV regime to be consistent with \gamray\ spectra from  Tycho's SNR as deduced from {\sl Fermi LAT} and {\sl VERITAS} observations 
\citep[][]{2017ApJ...836...23A}. 

The spectra shown in Fig.~\ref{fig:model1} do not show effects from \NL\ shock modification produced by a strong CR pressure gradient in the shock precursor \citep[e.g.,][]{JE91,MD2001,CBA2011}.
Shock modification modifies the spectral shape and normalization of CR spectra including the 
high-energy end. The effect has been studied extensively for some young 
SNRs like RX J1713.7-3946 and Cas A \citep[e.g.,][]{ESPB2012,Zirakashvili2014,Slane2015Err}. 
However, the polarization results we present here are 
qualitatively similar  with and without the \NL\ effects
so we have used the unmodified spectra shown in Fig.~\ref{fig:model1}.

\subsection{Model SNR geometry}
Tycho's SNR is among some 
young remnants having well developed small-scale structures in X-ray images. Here, we study the polarization properties of the magnetic turbulence believed associated with these structures in context of the new generation of X-ray polarimeters.
We have simulated synchrotron intensity and polarization maps 
of a Tycho-like SNR in X-ray energies taking into account stochastic
properties of the magnetic field. 
We assume a remnant radius of 3~pc, at a 
distance of 2.5~kpc, giving an angular 
radius of to 4 arcmin.\footnote{We note that $2.5$~kpc is at the low end of generally cited distances for Tycho's SNR which can be 3~kpc or a bit higher \citep[e.g.,][]{Slane2015Err,SatoHughes2017}.} 

\begin{figure}
\vskip12pt
\epsscale{1.0}
\plotone{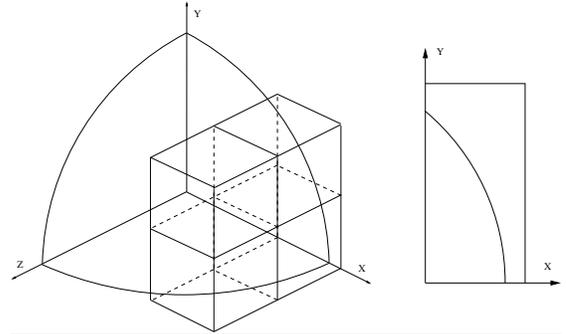}
\vspace{0.5 cm}
\plotone{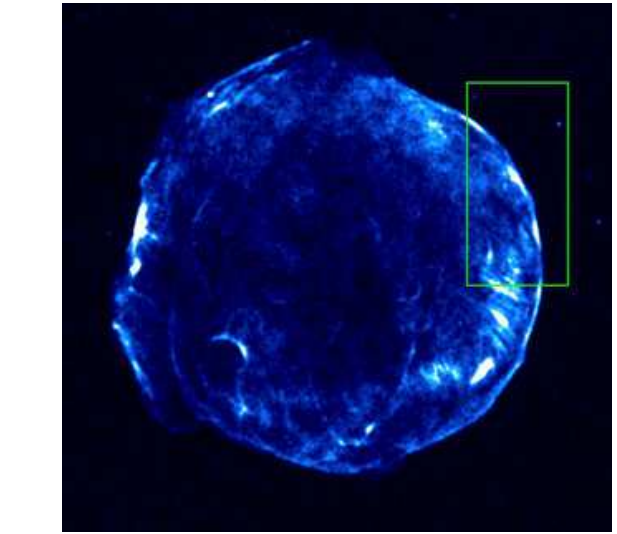}           
\caption{The model geometry is shown in the top left where 4 of the 8 simulation
boxes are shown.  One-eighth of the spherical SNR shock is indicated along with the coordinate
axes and, on the right, a schematic 2D image obtained after an integration over the line-of-sight. 
The Oz axis is directed to the observer. 
The bottom panel
shows the 6--8 keV \chandra\ image of Tycho's SNR with the location of
a $120\times240$ arcsec rectangular region which is equal in size
to our model region used  in simulations
in Section~3.
\label{fig:geometry}}
\end{figure}

We simulated magnetic fields in a rectangular box with dimensions  
$(16\times8\times4)\cdot10^{18}$~cm, consisting of eight cubes with sides  
$D=4\cdot 10^{18}$~cm.
As shown in Fig.~\ref{fig:geometry}, this box contains a part of the
SNR shell.
Our  minimal and maximal scales of turbulence are 
$\Lmin\sim2.5\cdot10^{15}$~cm and $\Lmax \sim1.2\cdot10^{18}$~cm. In angular units this is
$\Lmin \sim0.07$ arcsec and $\Lmax\sim30$ arcsec. We calculated
Stokes parameters of the synchrotron emission in this volume for different
models of isotropic and anisotropic magnetic turbulence both with $\sqrt{\Brms} =5.5\cdot10^{-5}G$.

After integration over the line-of-sight 
(OZ axis in Fig.~\ref{fig:geometry})
we obtained 2D intensity and polarization maps of the synchrotron
radiation. We used 2000 pixels along the line-of-sight and 100x100
pixels in a transverse plane in each cubic box. On the right panel
of Fig.~\ref{fig:geometry} the $120\times240$ arcsec simulated rectangular
region is shown superimposed on the Tycho's SNR \chandra\ image. 

We simulated  the axially symmetric magnetic turbulence with an axis of symmetry directed along the Ox axis
while the real symmetry of the system is spherical. Our simulation results are precise at the part of the SNR 
shell normal to the Ox axis (y=0) 
and accurate enough for our simulation maps in nearby regions.

\section{Magnetic turbulence and SNR synchrotron images}
\subsection{Ideal observations}
Here we discuss `ideal' observations, i.e., ones with long enough exposures so  all statistically uncertainties are negligible for
every pixel.
We do not, however, assume infinite angular resolution.
This ideal case 
allows us to identify synchrotron emission properties 
of young SNRs and serves as a starting point for simulations of real
synchrotron images taking into account polarimeter sensitivity, effective
area, limited exposure, and Poisson photon statistic. These realistic simulations
are discussed in Section~\ref{sec:real}.

Based on the model described above we simulated synchrotron maps for
Stokes parameters $I$, $U$, and $Q$ assuming an 
angular resolution of 1.2 arcsec. 
This allows us to obtain maps of the intensity, polarization degree,
and polarization angle with the same or worse resolution of the
forthcoming \ixpe\ X-ray polarimeter \citep[see][]{Weisskopf_2016}. 

\begin{figure}           
\vskip12pt
\epsscale{0.57}
\begin{center}{Isotropic}\end{center}
\begin{tabular}{cc}
$\delta=1$ & $\delta=5/3$ \\
{\plotone{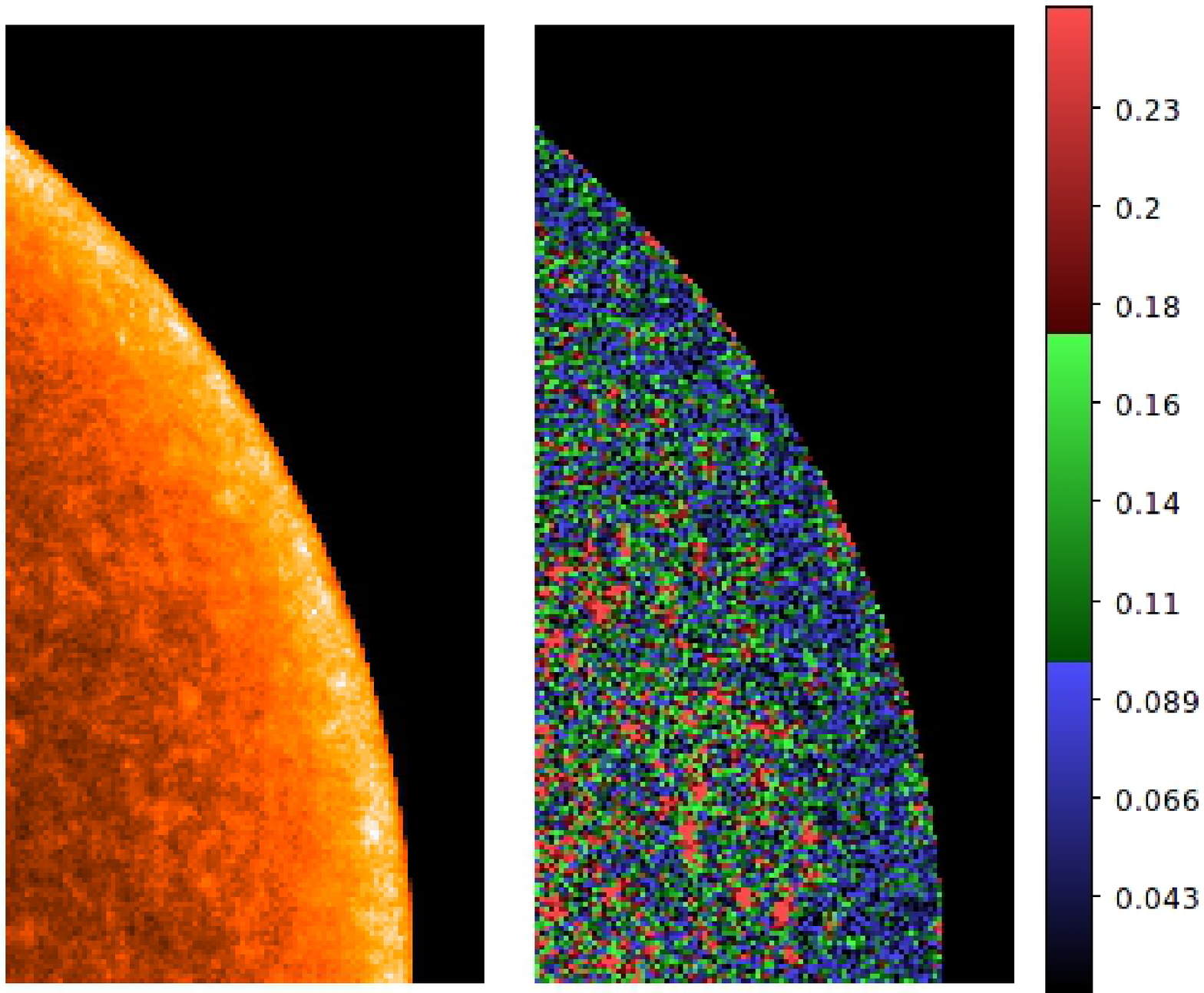}} & {\plotone{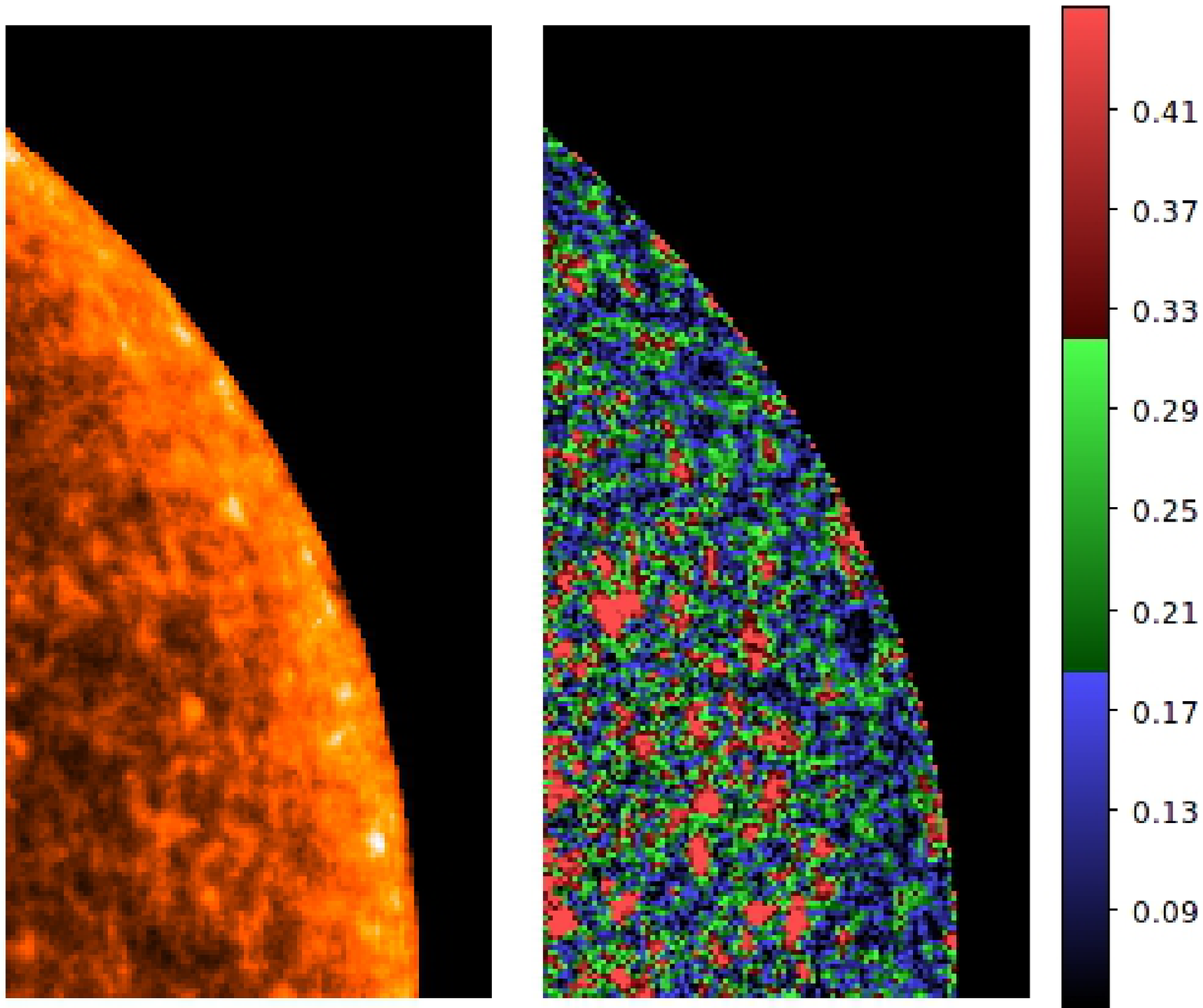}} \\
{\plotone{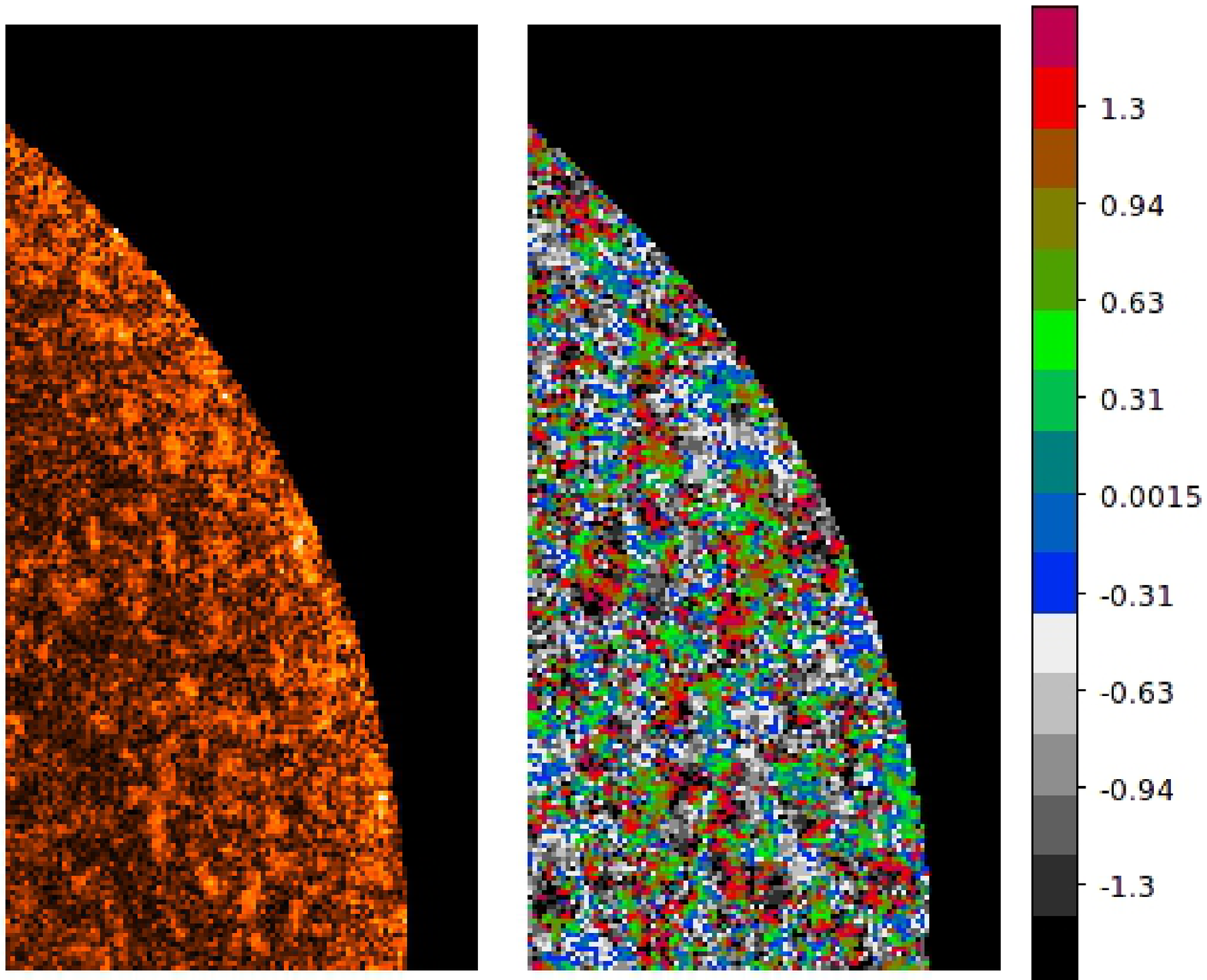}} & {\plotone{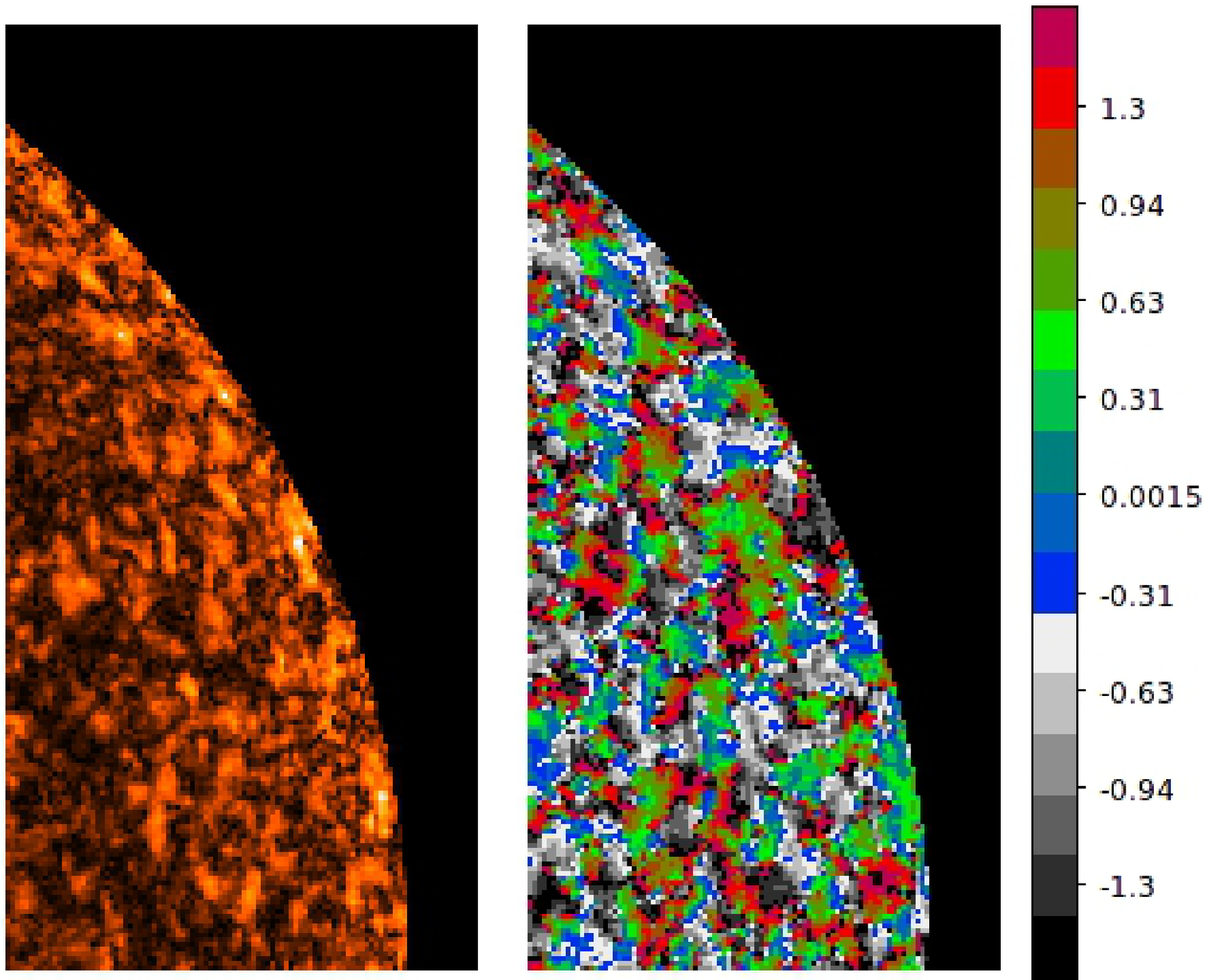}} \\
\end{tabular}
\begin{center}{Anisotropic}\end{center}
\begin{tabular}{cc}
$q=5$ & $q=20$ \\
{\plotone{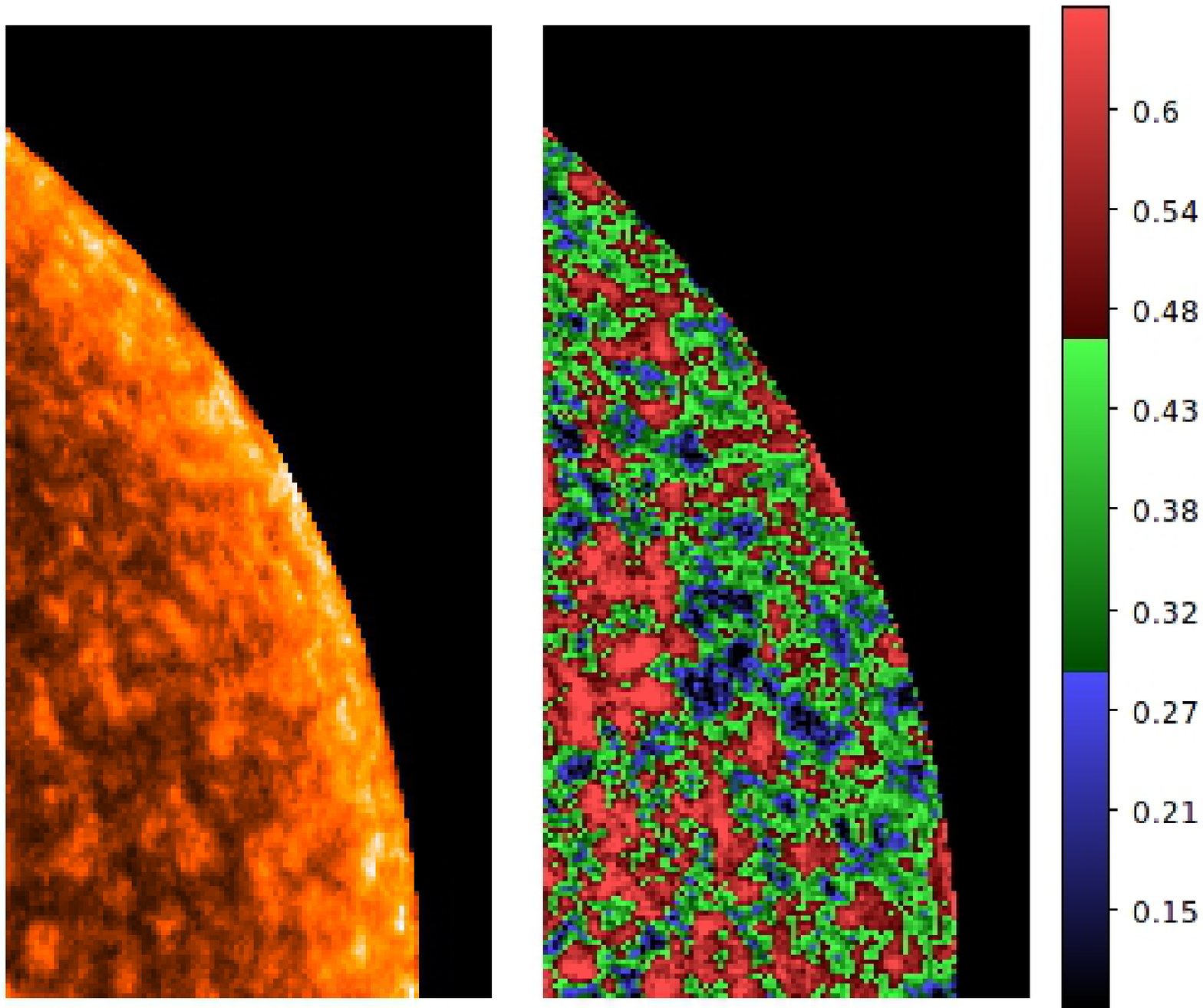}} & {\plotone{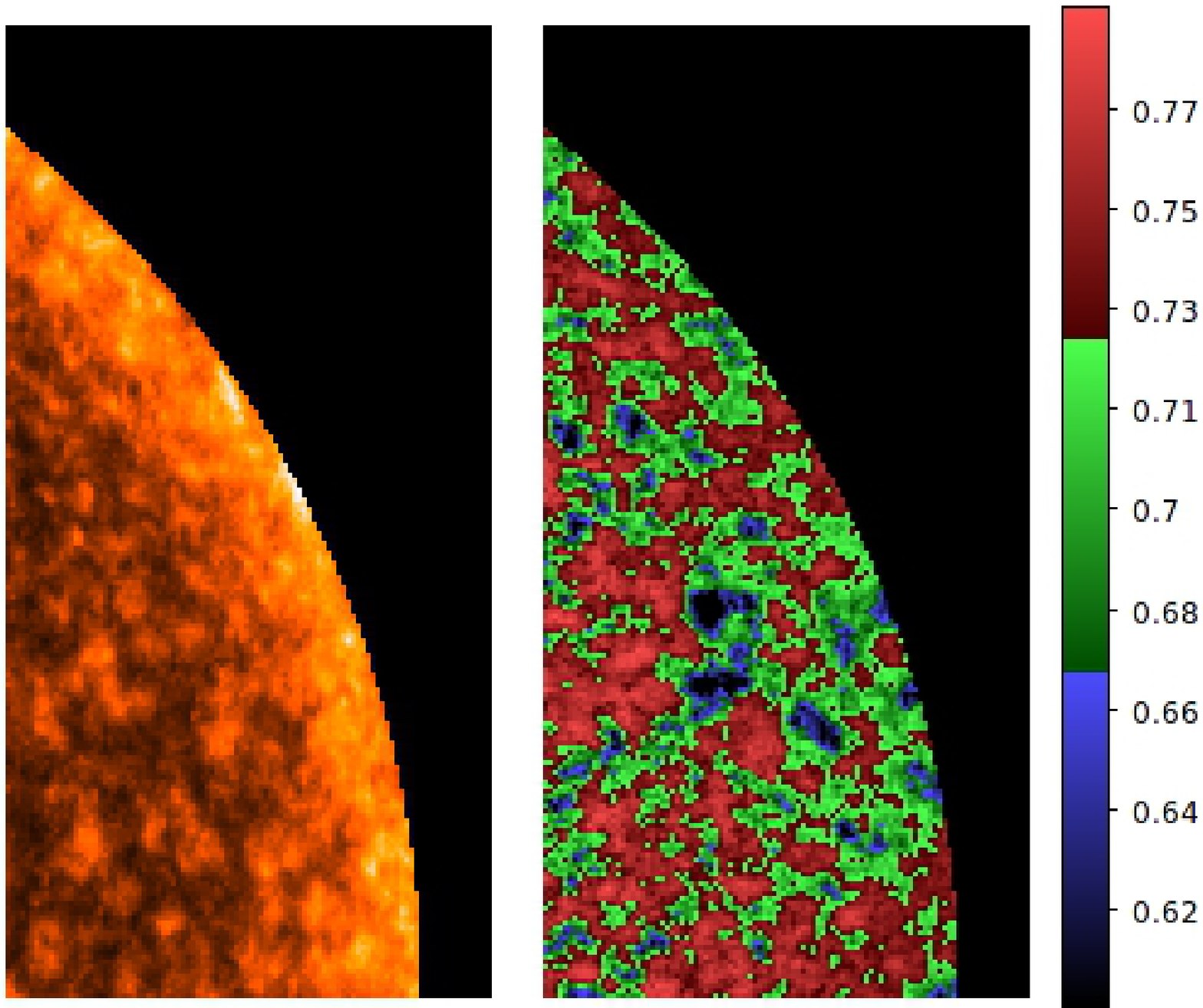}} \\
{\plotone{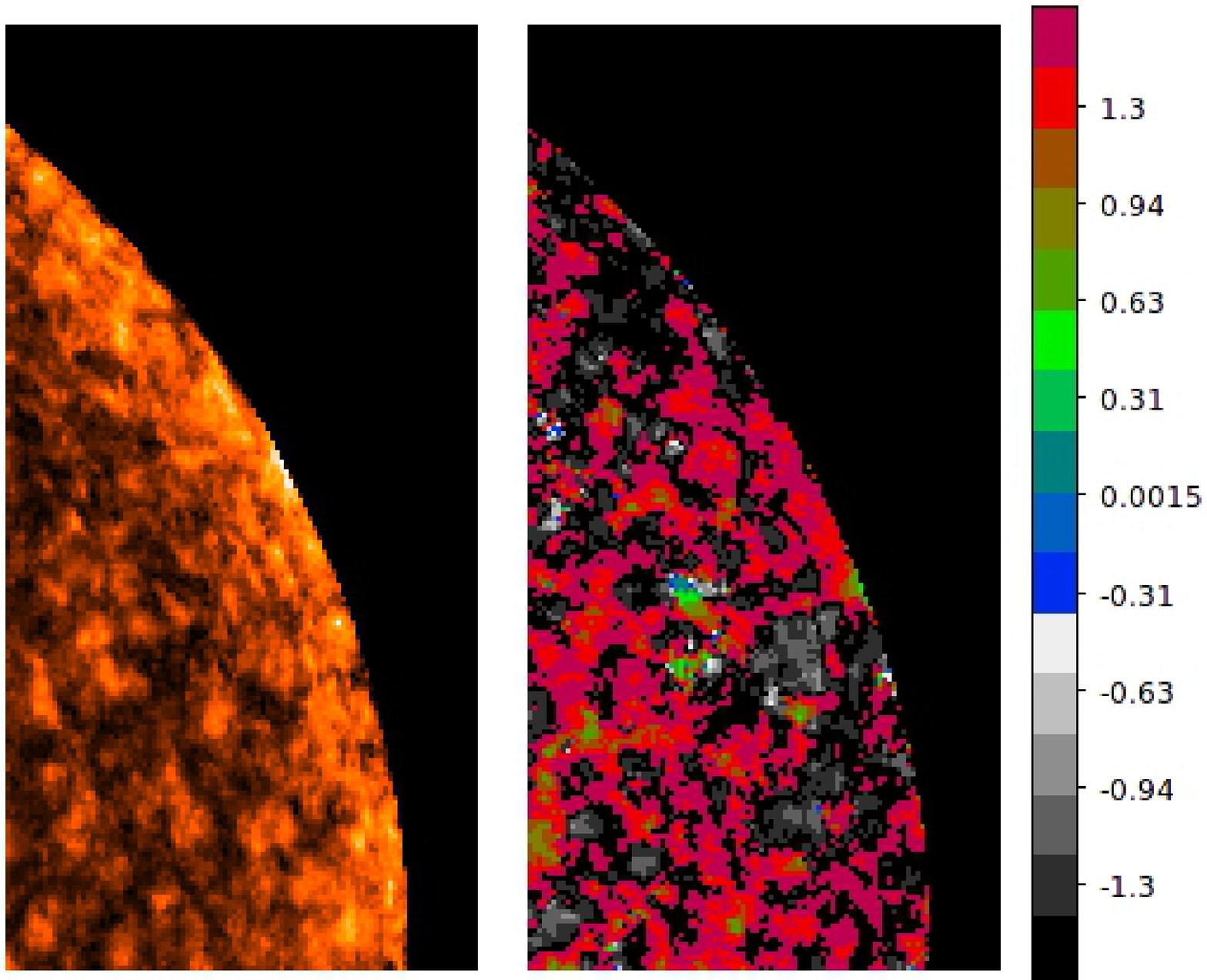}} & {\plotone{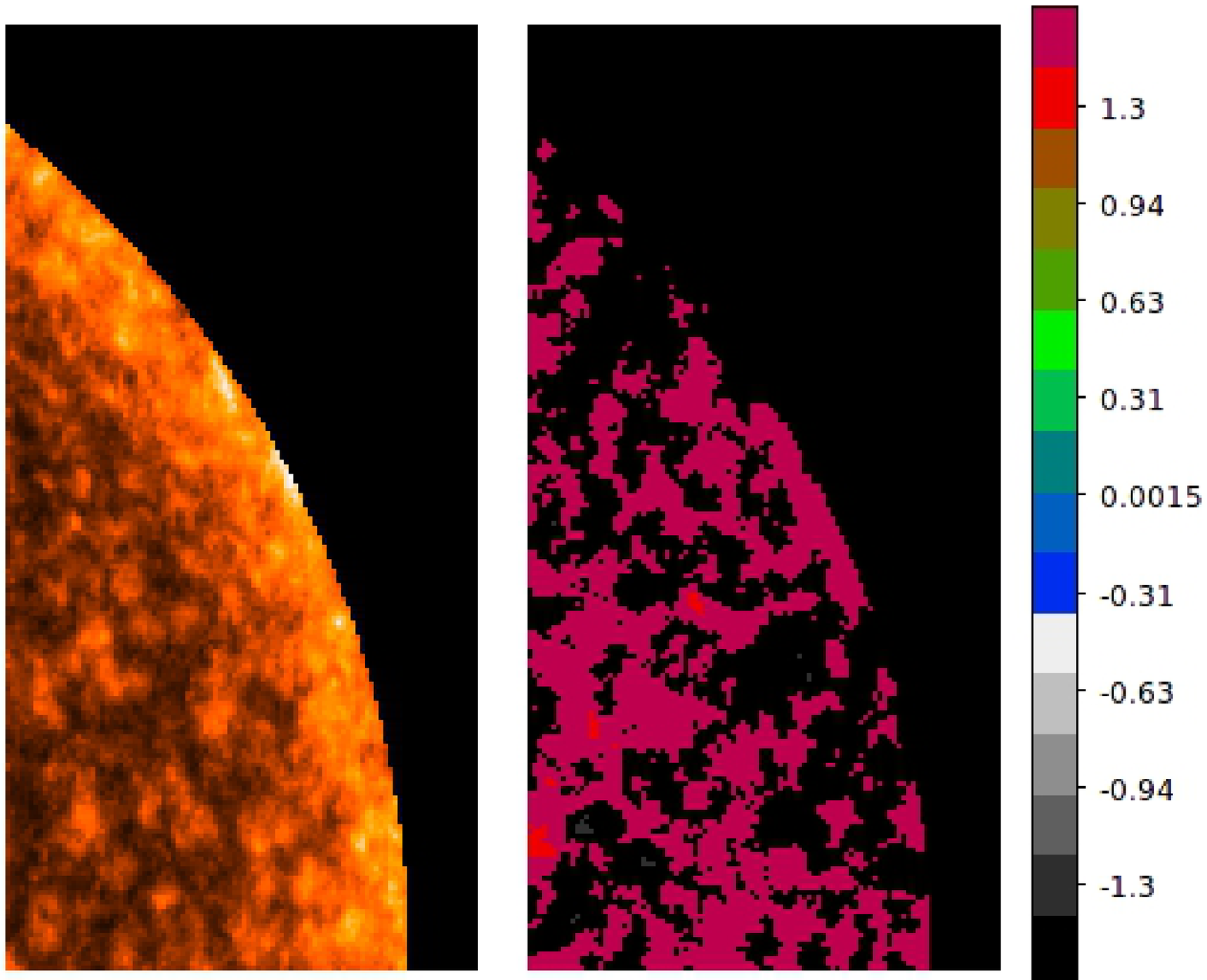}} \\
\end{tabular}          
\caption{Model synchrotron images with 1.2'' angular resolution.
In each of the 4 sections, the total emission is in the upper left, the polarized
emission is in the lower left, the polarization degree is in the upper
right and the polarization angle (in radians) is in the lower right.
Angles are measured from the Oy axis. The left and right panels
in the first row, respectively, show isotropic turbulence examples for 
turbulence power spectra indices $\delta$=1 and 5/3. The left and
right panels in the bottom row, respectively, show anisotropic
turbulence cases with $\delta$=5/3 and $q=5$ and 20.
\label{fig:1.2_isotr_and_anisotr}}
\end{figure}

We simulated cases with isotropic and anisotropic magnetic
turbulence.
For the isotropic case, we used magnetic field fluctuations with power spectrum indices $\delta=5/3$ and  $\delta=1$.
In the anisotropic case, we consider both turbulence produced
due to shock compression and due to anisotropic cascading. 
In the shock compression case, we consider magnetic turbulence with power
spectrum index $\delta=5/3$ and strength of anisotropy $q=5$ and $20$.
Radiation maps obtained in these cases for a photon energy of 
5~keV are
shown in Fig.~\ref{fig:1.2_isotr_and_anisotr} with 1.2~arcsec
angular resolution.
The dependence on the spectral index is clearly
seen in the relative sizes of magnetic structures in the high resolution
intensity and polarization images for different values of $\delta$.
It is also seen that stronger anisotropy (greater $q$) in shock
compression turbulence results in stronger polarization 
with mostly a radial polarization direction.

\begin{figure}    
\vskip12pt
\epsscale{0.57}
\begin{center}{Isotropic}\end{center}
\begin{tabular}{cc}
$\delta=1$ & $\delta=5/3$ \\
{\plotone{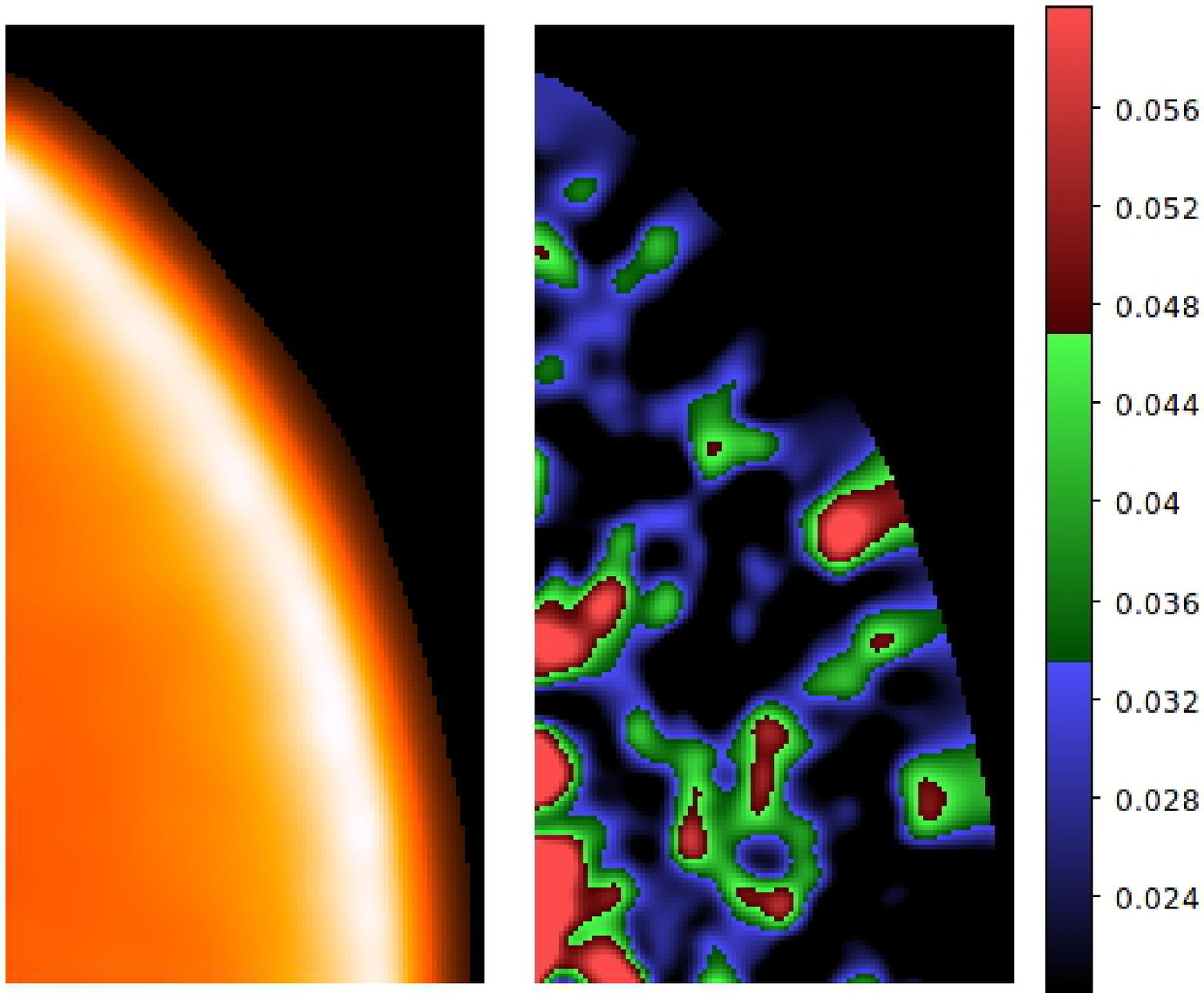}} & {\plotone{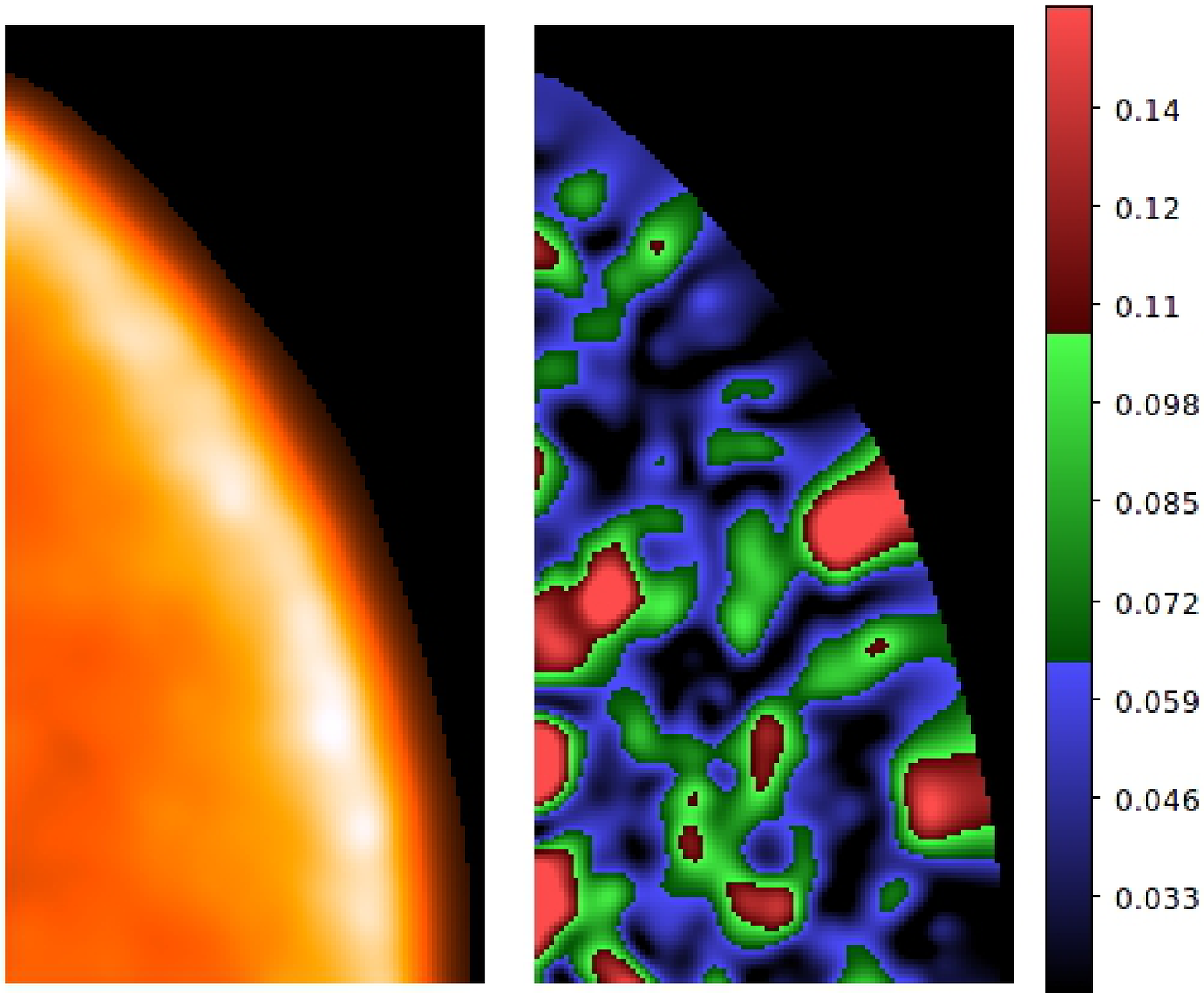}} \\
{\plotone{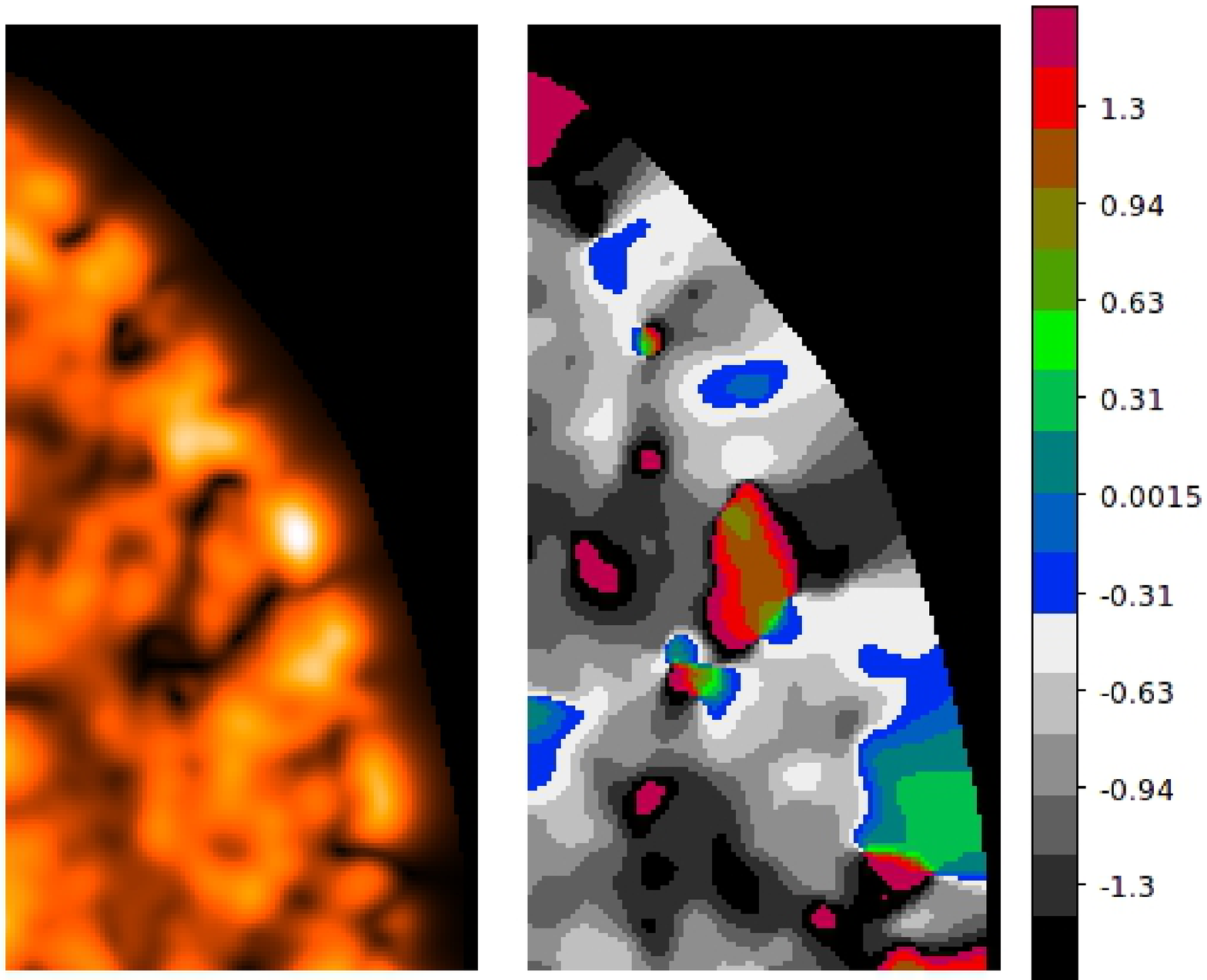}} & {\plotone{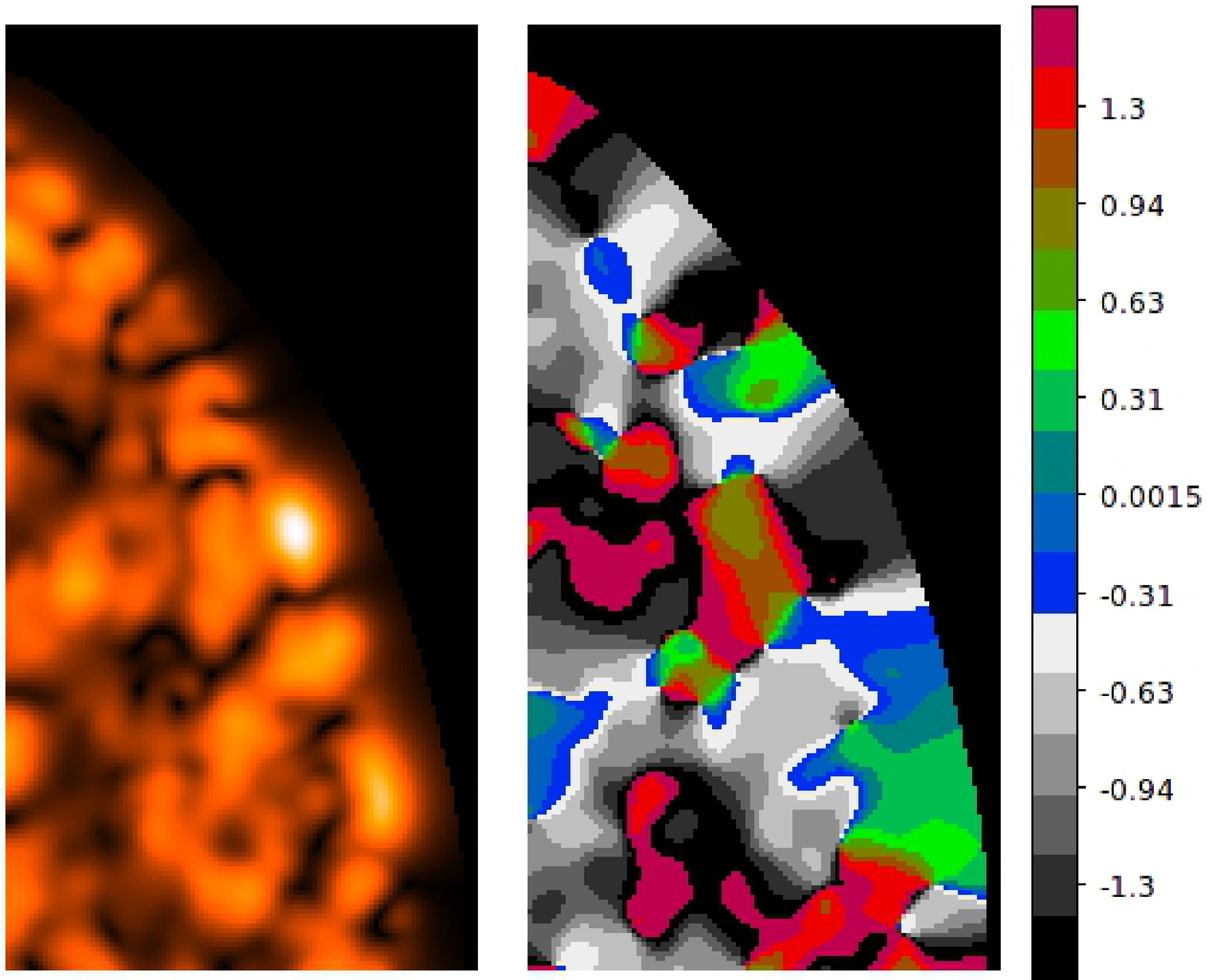}} \\
\end{tabular}
\begin{center}{Anisotropic}\end{center}
\begin{tabular}{cc}
$q=5$ & $q=20$ \\
{\plotone{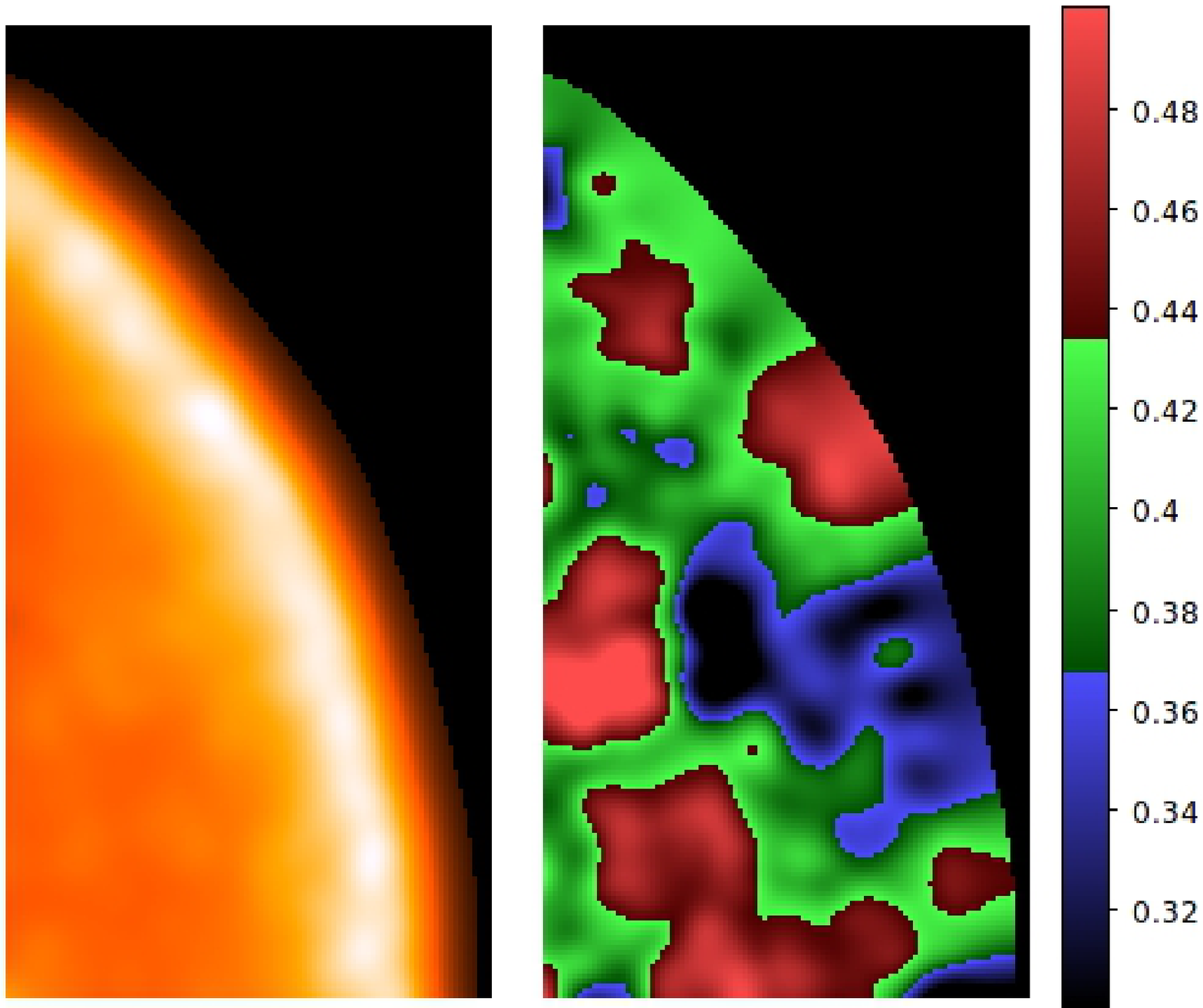}} & {\plotone{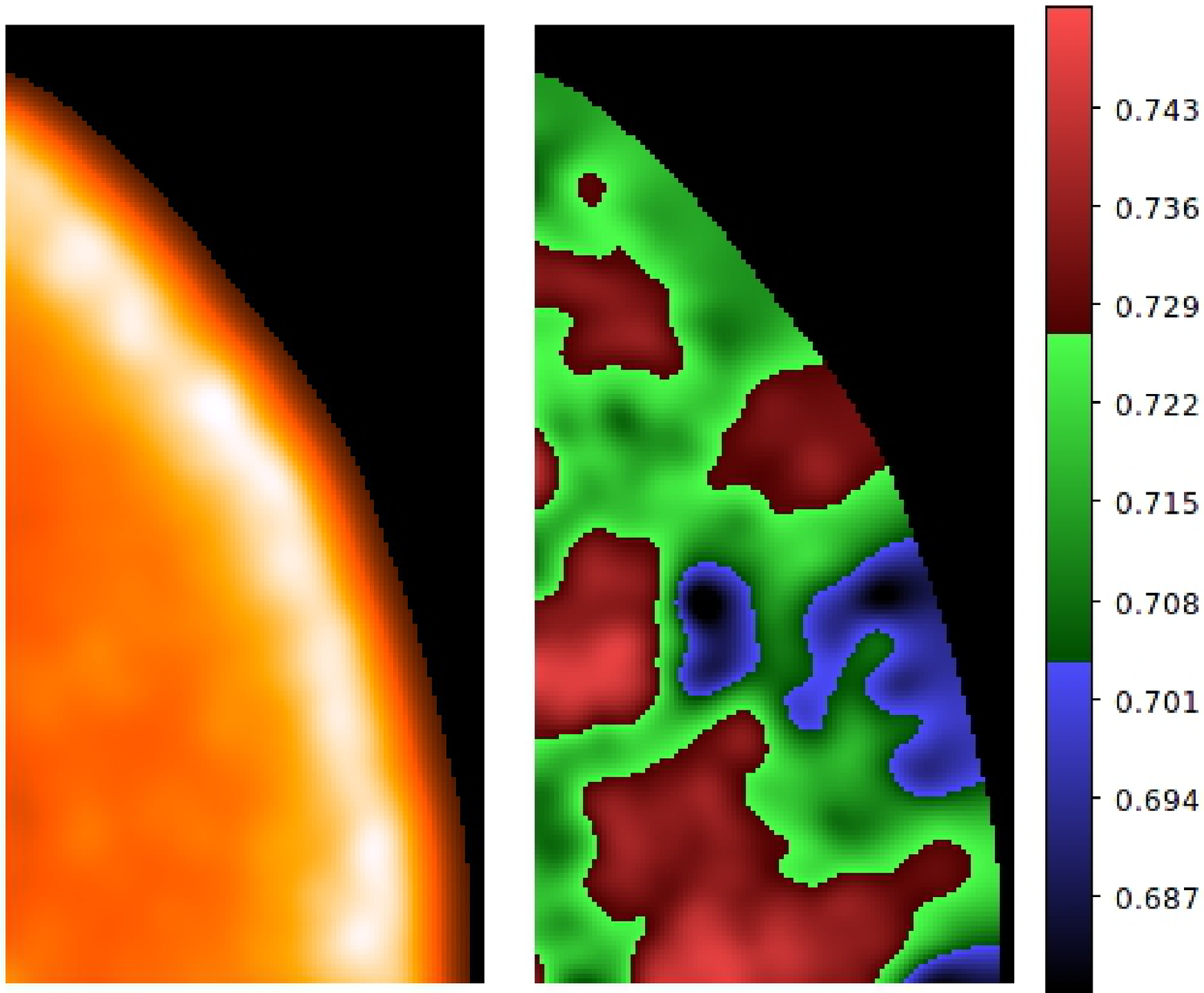}} \\
{\plotone{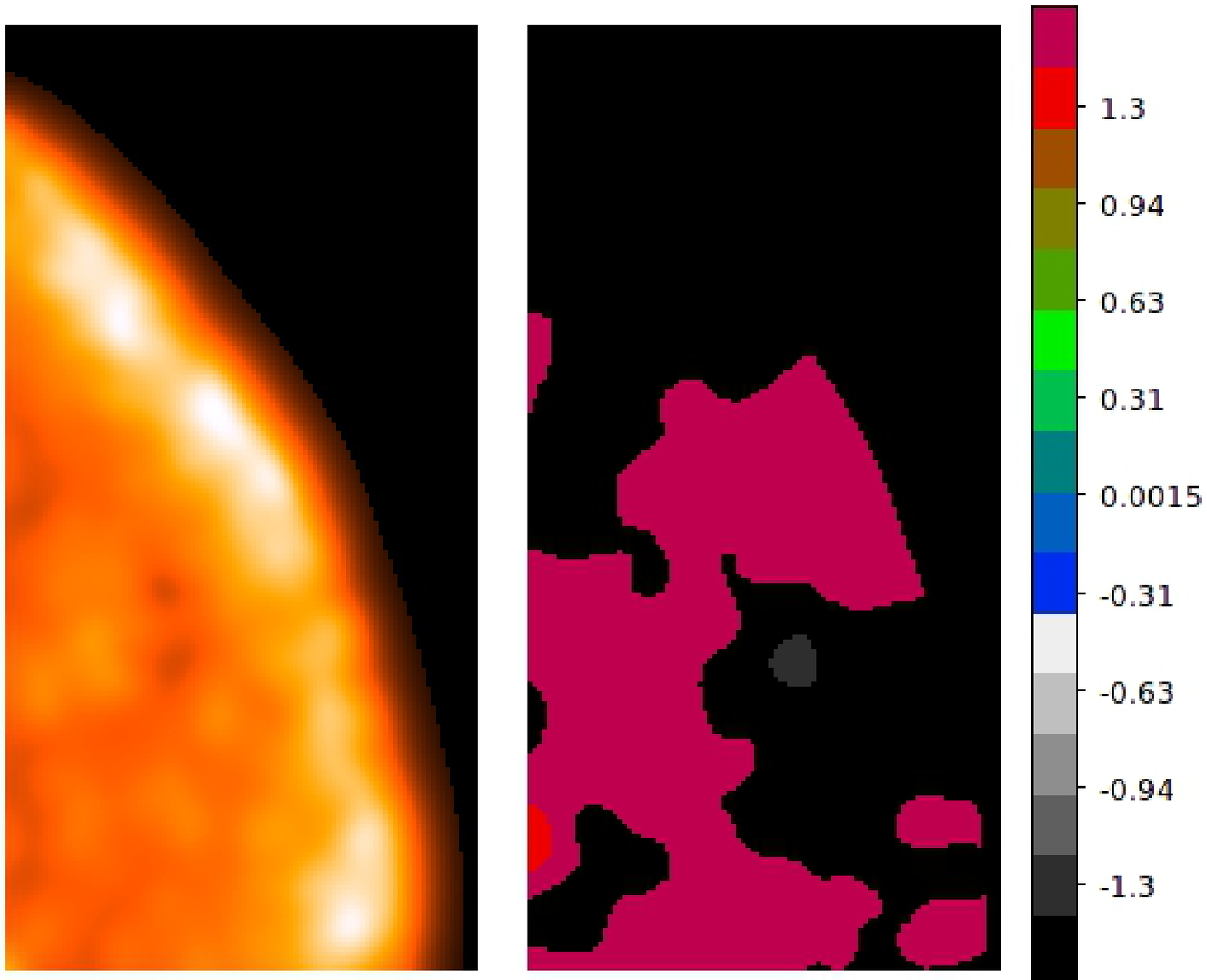}} & {\plotone{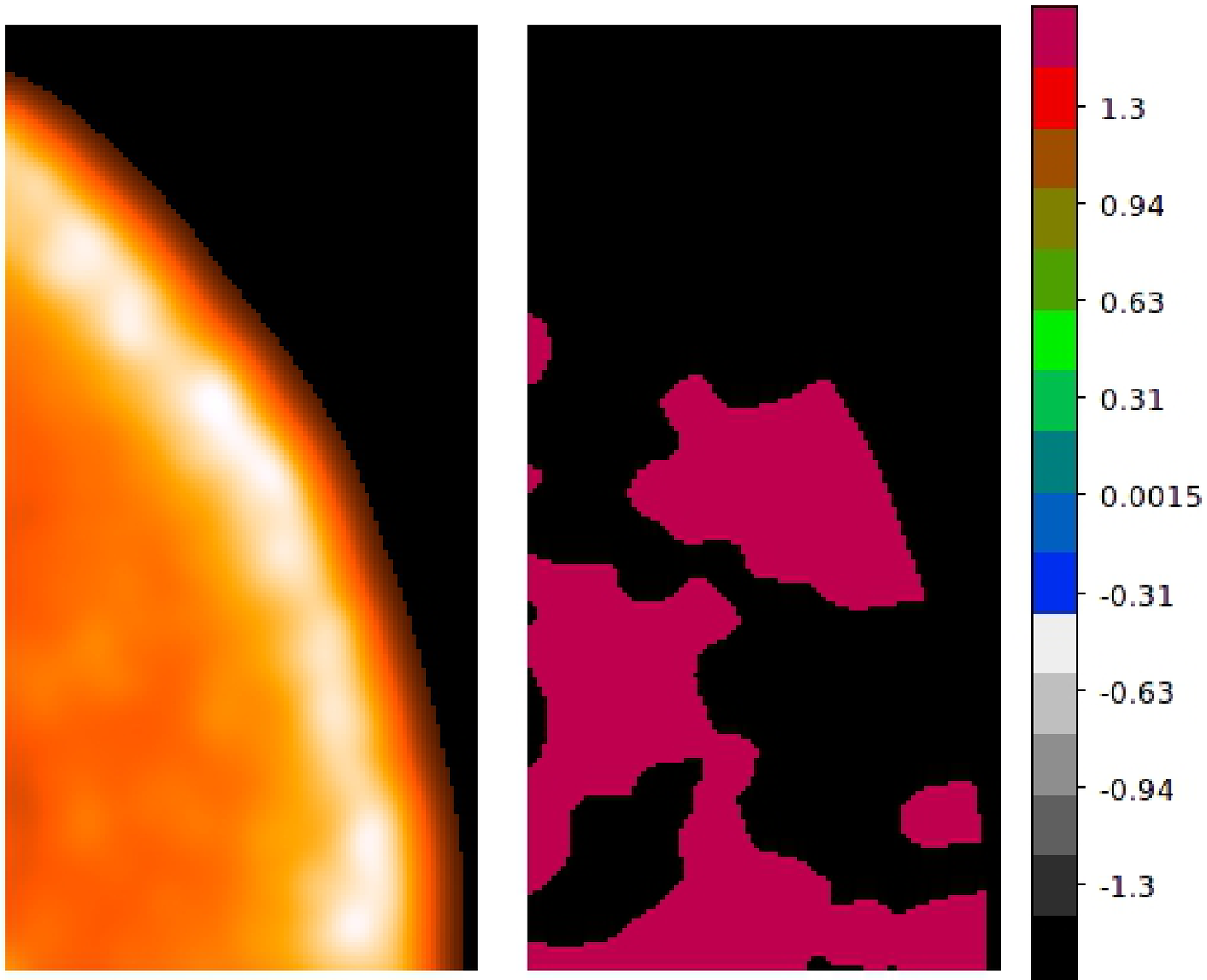}} \\
\end{tabular}           
\caption{Model synchrotron images obtained after convolution 
with the \ixpe\ PSF are shown. In each of the 4 sections, the total emission is in the upper left,
the polarized emission is in the lower left, the polarization degree
is in the upper right, and the polarization angle (in radians) is in
the lower right. Angles are measured from the Oy axis. 
The left and
right panels in the top row, respectively, show isotropic turbulence
examples  for turbulence power spectra indices $\delta=1$ and  5/3. The
left and right panels in the bottom row, respectively, show  anisotropic
turbulence examples with $\delta=5/3$ and $q=5$ and 20.
\label{fig:XIPE_PSF_isotr_and_anisotr}}
\end{figure}

This agrees well with the fact that the polarization in highly anisotropic turbulence approaches the  maximum theoretical limit of polarization in a homogeneous magnetic field.
The main difference comes about because, for a homogeneous magnetic
field, the total emission from the entire
SNR is highly polarized. For anisotropic turbulence, only small scale
structures are highly polarized and 
the total emission  is not significantly polarized because contributions
 from different parts of the remnant
cancel each other  on average. 

In Fig.~\ref{fig:XIPE_PSF_isotr_and_anisotr} we show maps of the quantities from 
Fig.~\ref{fig:1.2_isotr_and_anisotr}
with the expected angular resolution of the \ixpe\ polarimeter as a convolution with 
PSF taken from \citep{Fabiani_2014}. 
The work of  \citet{Fabiani_2014} describes the PSF of the XIPE polarimeter and
we assume this is  similar to the PSF of \ixpe. 
The PSF was truncated to a circle with 36 arcsec radius in order
to decrease the influence of border effects. 
In the maps shown in Fig.~\ref{fig:XIPE_PSF_isotr_and_anisotr}, the  
\ixpe\ PSF
width exceeds the smallest  scale structures
so, after convolution, these  structures will  become
much less prominent. 
Nevertheless, there is a  difference in polarization between the homogeneous and anisotropic cases after convolution so future X-ray polarimeters
could give quantitative information on  the level of anisotropic magnetic turbulence in SNRs.

To model turbulent magnetic field produced in a cascading process
we used power spectra of weak turbulence cascade as described
in Section~2.1. 
The stochastic field properties correspond
to  $q=1.05$. If no large-scale homogeneous
field is present, this stochastic field
alone can be used to  model strong turbulence cascade. 
Weak turbulence can  be modeled with the addition of a  large-scale homogeneous field in which case the strength of turbulence
depends on the ratio between $\Bhom^2$ and $\BturSq$.
In all of our examples, we fix the value of
$\langle {\bf B}^2 \rangle = 
\langle ({\bf \Bhom} +  {\bf \Btur})^2 \rangle = {\rm const}$, 
an important factor for electron diffusion and energy loses. 
We
chose a direction of the large-scale magnetic field, which is also
the symmetry axis, along the axis Ox. In the model
geometry (Fig.~\ref{fig:geometry}), this direction mimics a radially
directed large-scale magnetic field.

\begin{figure}         
\vskip12pt
\epsscale{0.57}
\begin{center}{Strong cascade}\end{center}
\begin{tabular}{cc}
high angular resolution & \ixpe\ angular resolution \\
{\plotone{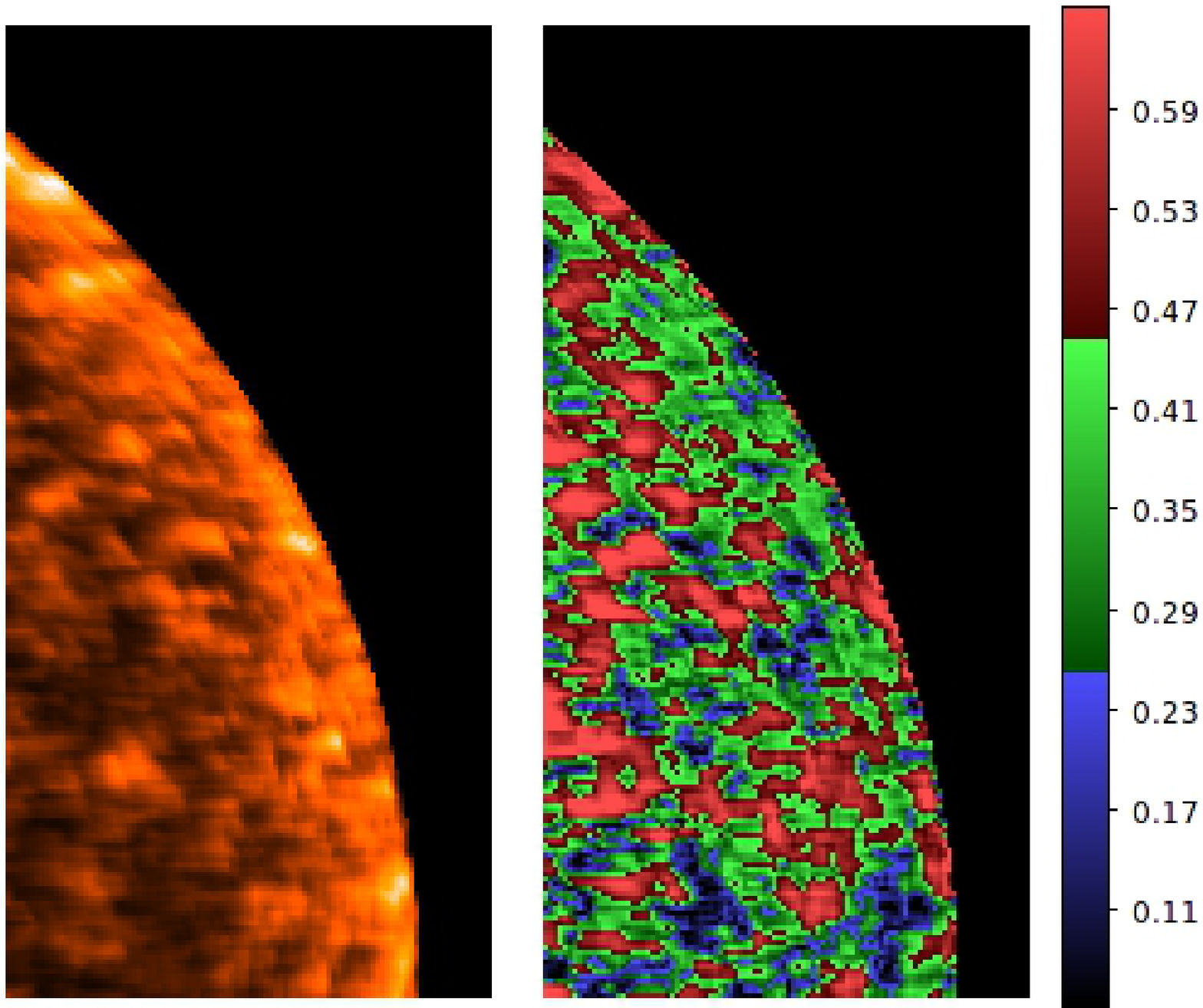}} & {\plotone{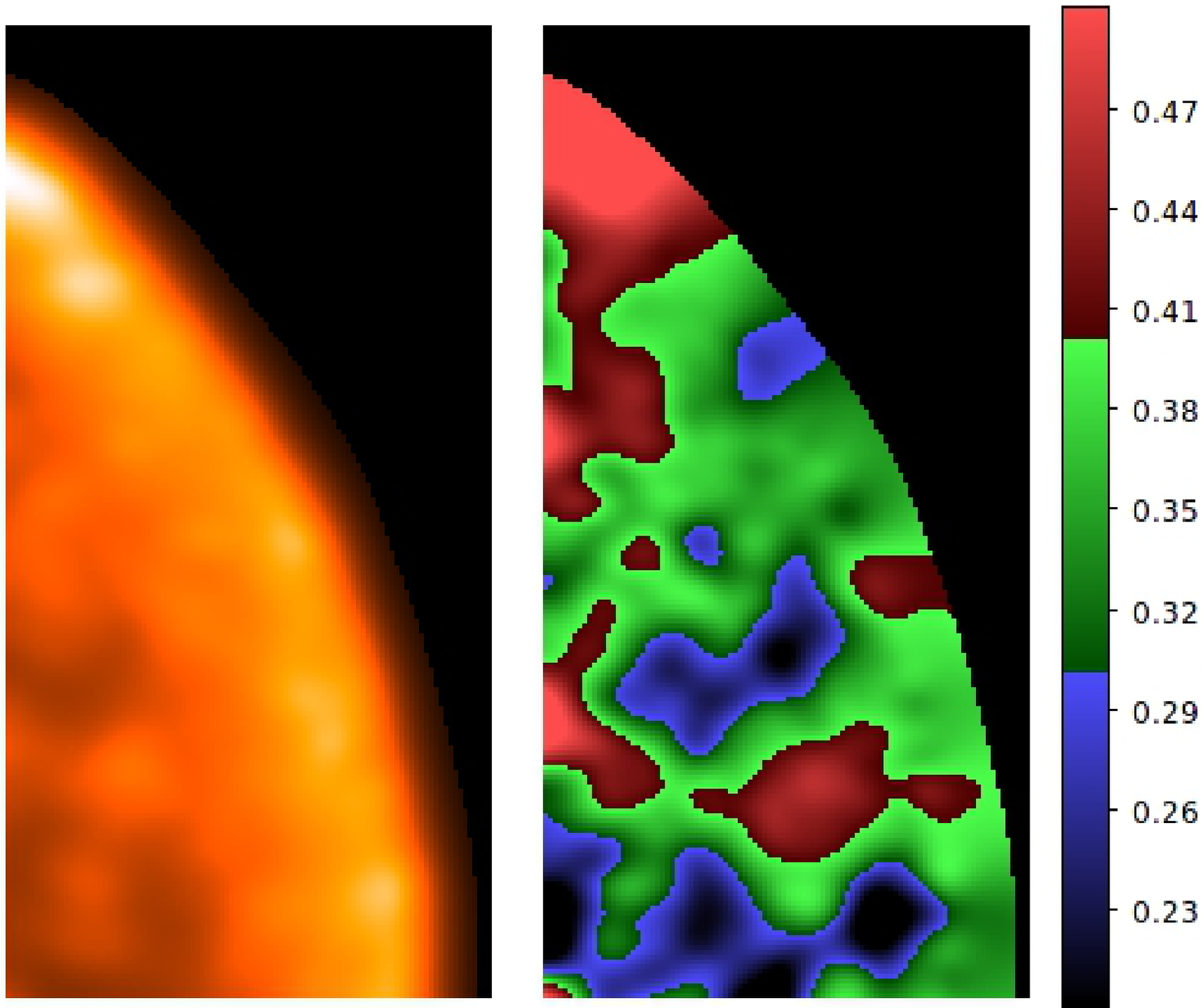}} \\
{\plotone{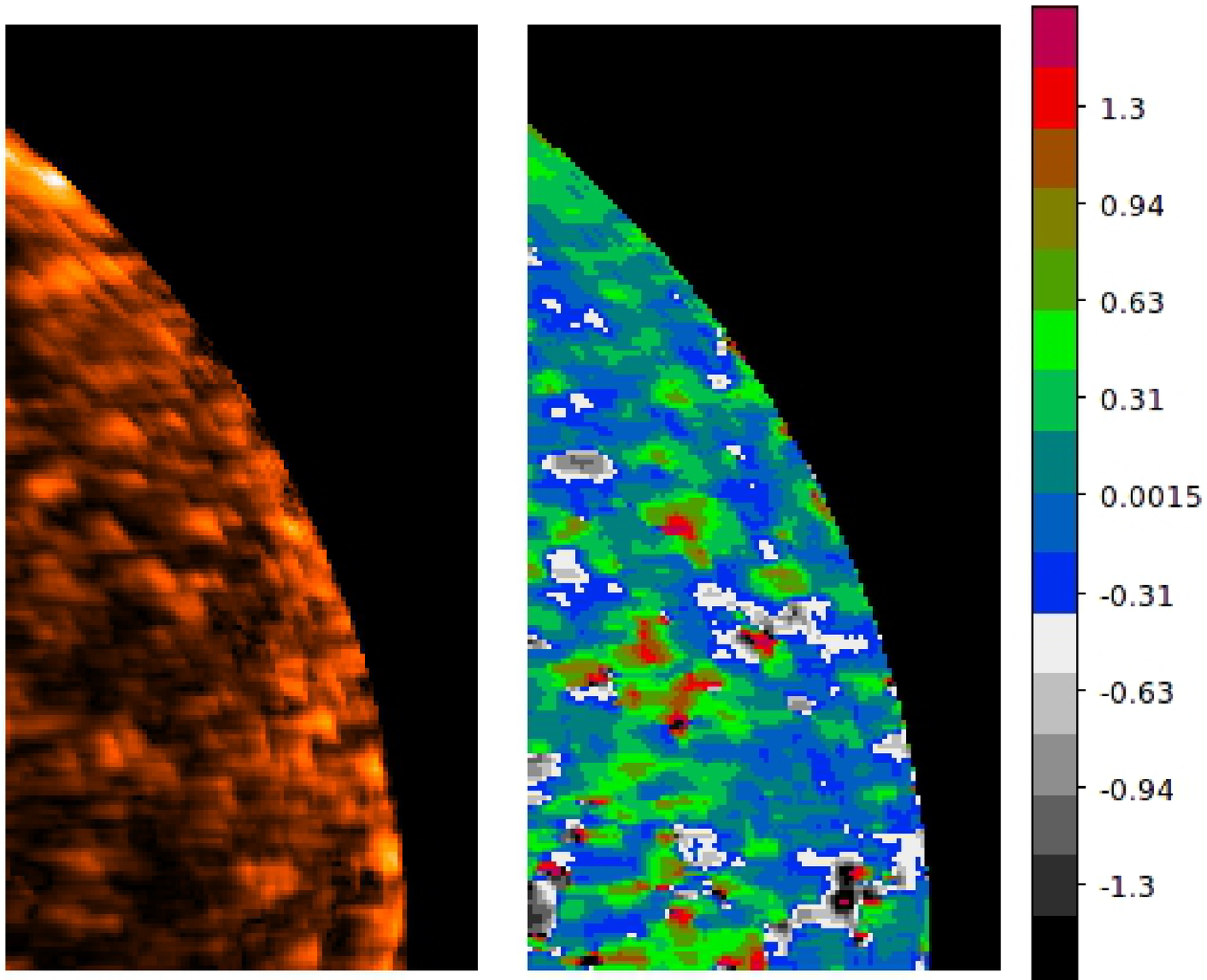}} & {\plotone{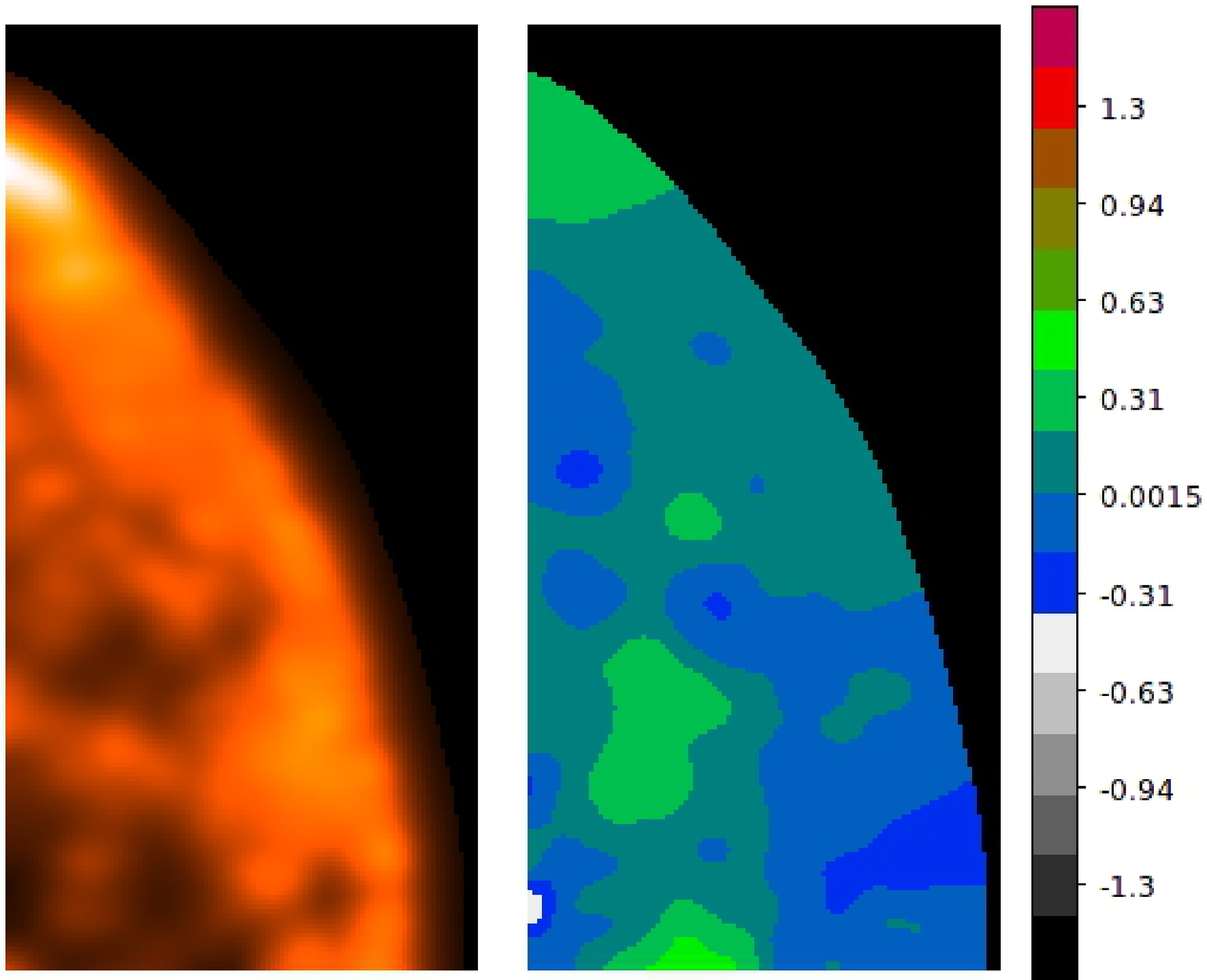}} \\
\end{tabular}
\begin{center}{Weak cascade}\end{center}
\begin{tabular}{cc}
high angular resolution & \ixpe\ angular resolution \\
{\plotone{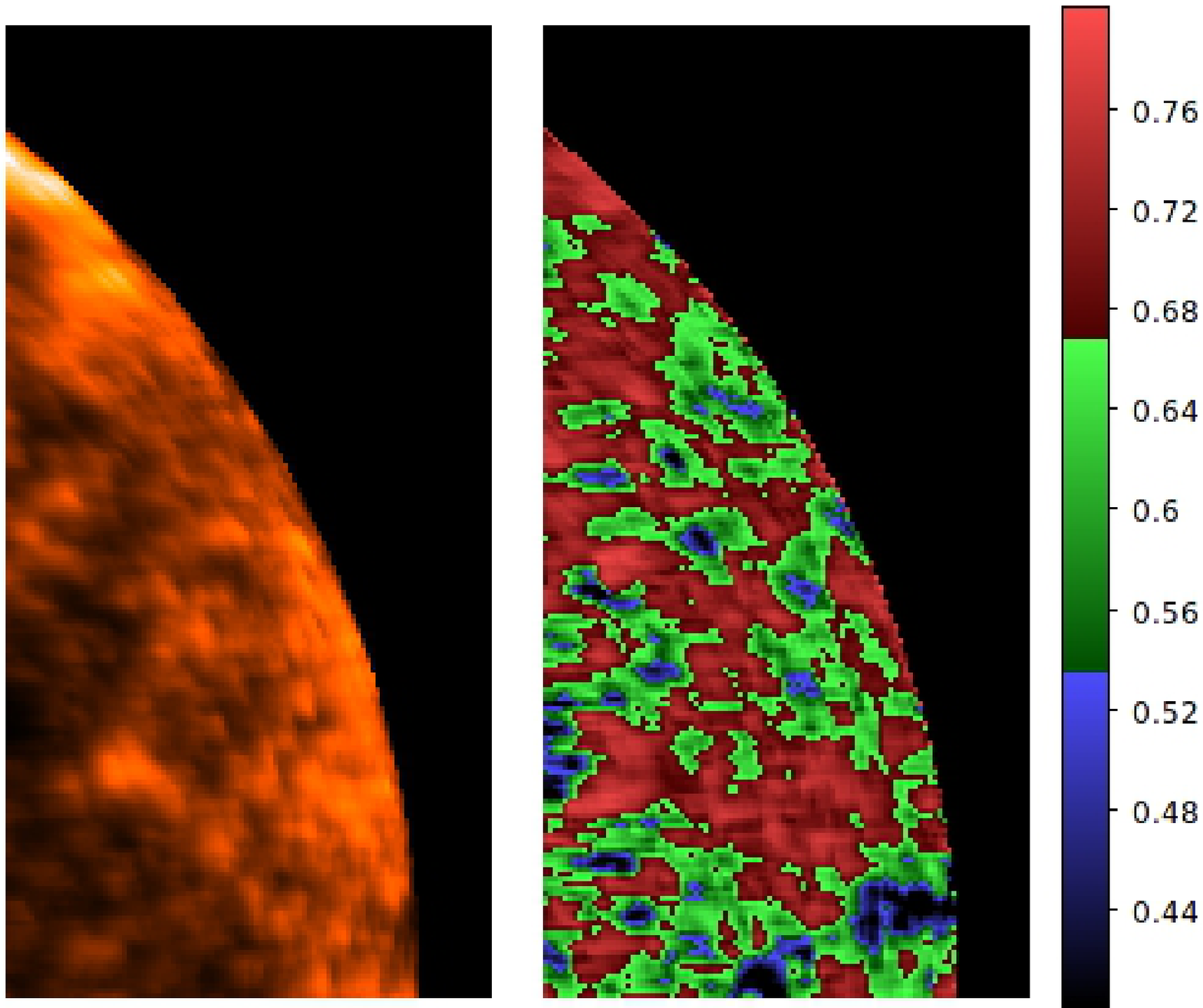}} & {\plotone{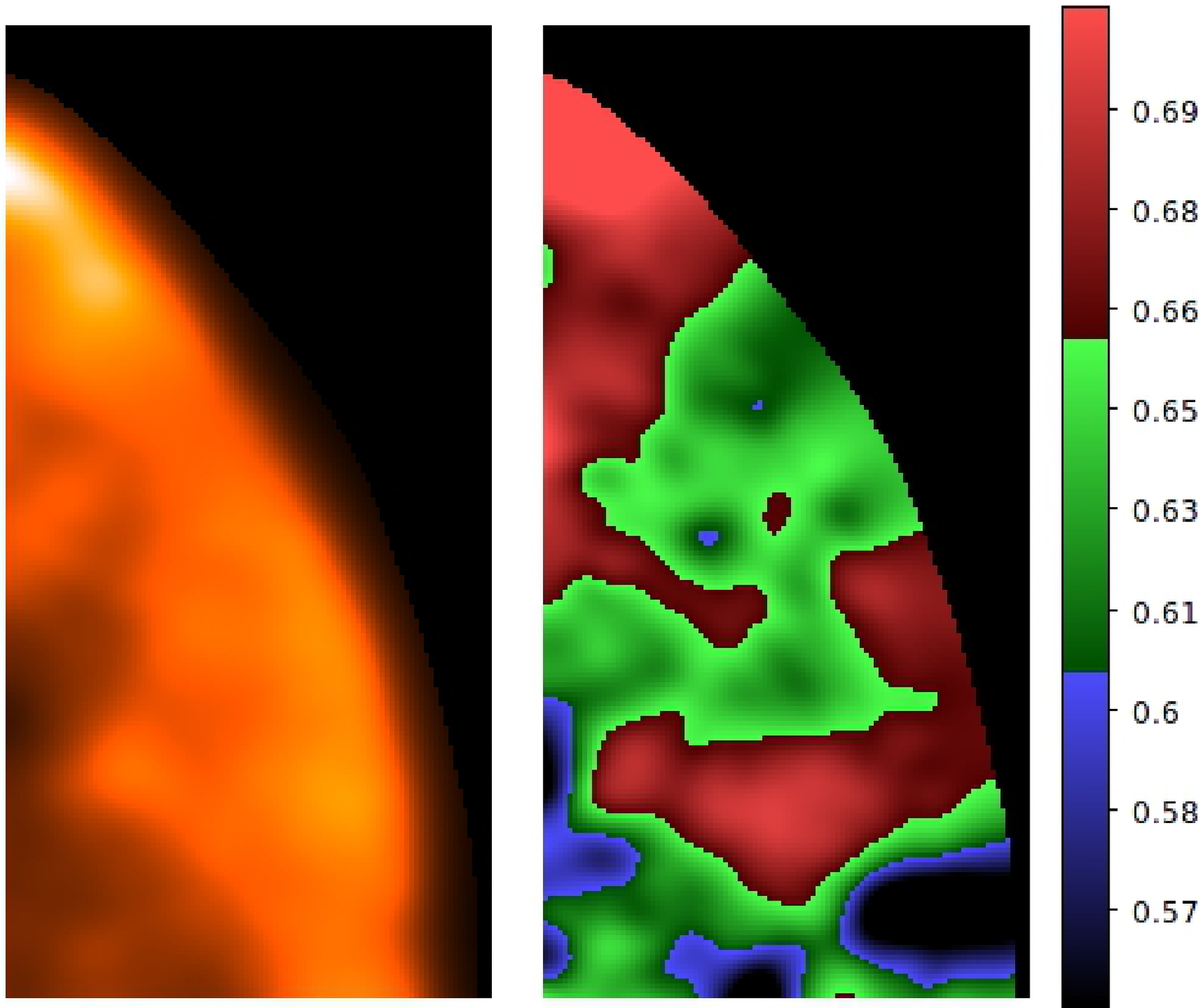}} \\
{\plotone{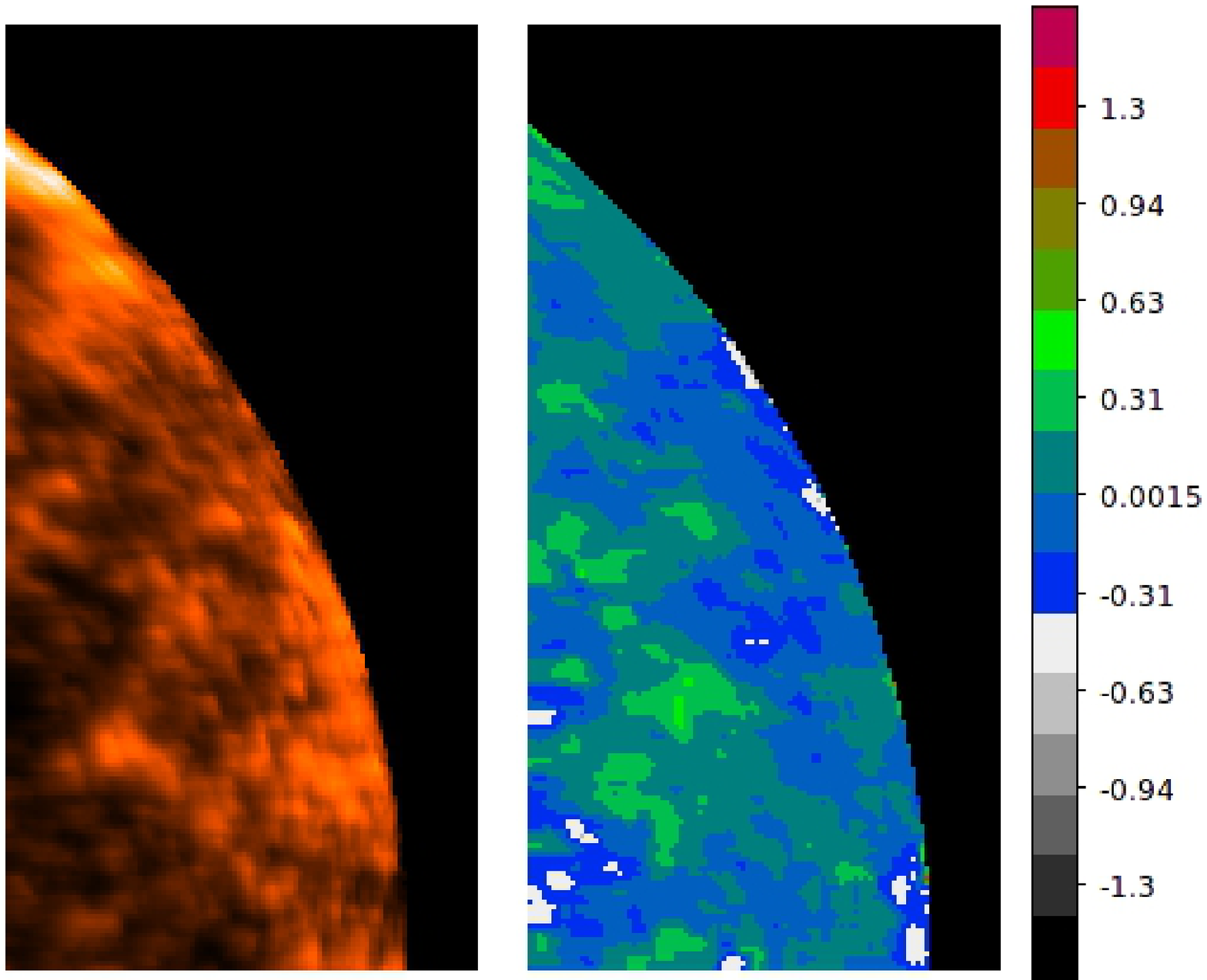}} & {\plotone{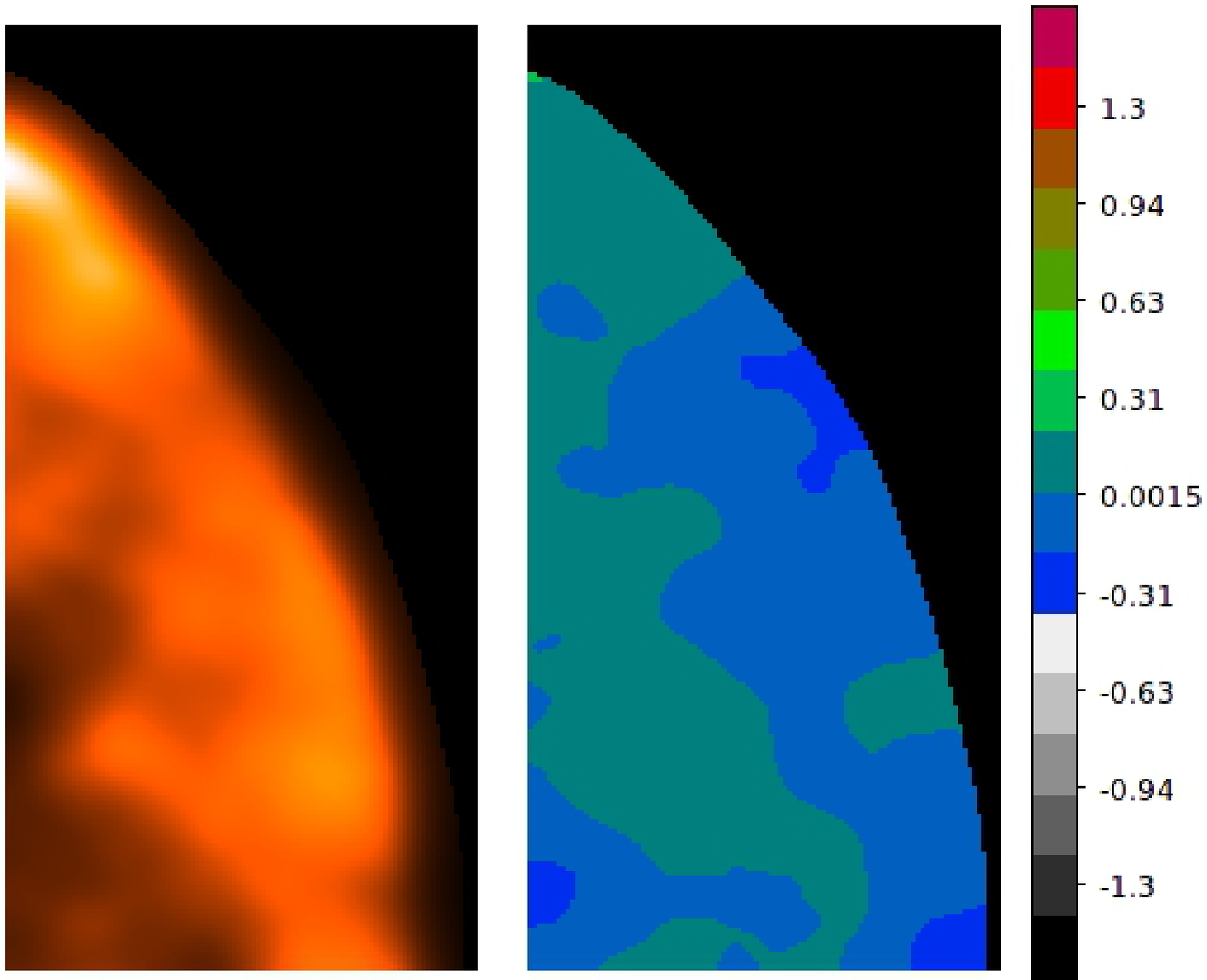}} \\
\end{tabular}          
\caption{These images were generated  with turbulence produced in an
anisotropic cascade process. In each panel the total emission is in
the upper left, the polarized emission is in the lower left, the polarization
degree is in the upper right, and the polarization angle (in radians)
is in the lower right. Angles are measured from the Oy axis. Images
with angular resolution 1.2\char`\"{} are in the left column and images
obtained after convolution with \ixpe\ PSF are in the right column.
The first row shows the case of anisotropic turbulence produced in
the strong anisotropic cascade model with $q=1.05$. The second row shows
the case of anisotropic turbulence produced in the weak anisotropic cascade
model (see text).
\label{fig:cascade_image}}
\end{figure}

Fig.~\ref{fig:cascade_image} shows synchrotron images for these turbulent fields. 
It can be seen that, while radiation is more polarized for  weaker turbulence, there 
remains significant polarization for strong turbulence.
In both cases, the most likely direction of polarization is along the
Oy axis, i.e., tangent to the shock front. 
This direction differs from the one for turbulence produced by shock compression. This result remains
after convolution with the \ixpe\ PSF making it possible to distinguish
anisotropic cascade turbulence from shock compressed
turbulence.

\subsection{Simulation of realistic \ixpe\ observations} \label{sec:real}
Realistic simulations of the X-ray band should take into account
the detector sensitivity and the Poisson statistics  of the incoming photons.
Supernova remnants are extended objects so even if the entire remnant is a strong source,
spatially resolved parts are likely to be rather weak sources. 
In \citet{Weisskopf_2010} it is stated that detection of polarization requires much 
better statistics than needed for spectral analysis. With this in mind, we explore the 
requirements to detect and study the polarization properties of Tycho's SNR by 
the forthcoming \ixpe\ polarimeter.

\begin{figure}    
\vskip12pt
\epsscale{1.2}
\begin{center}{Anisotropic turbulence with $q=5$}\end{center}
\includegraphics[trim={0 2.7cm 0  1.7cm},clip,width=0.5\textwidth]{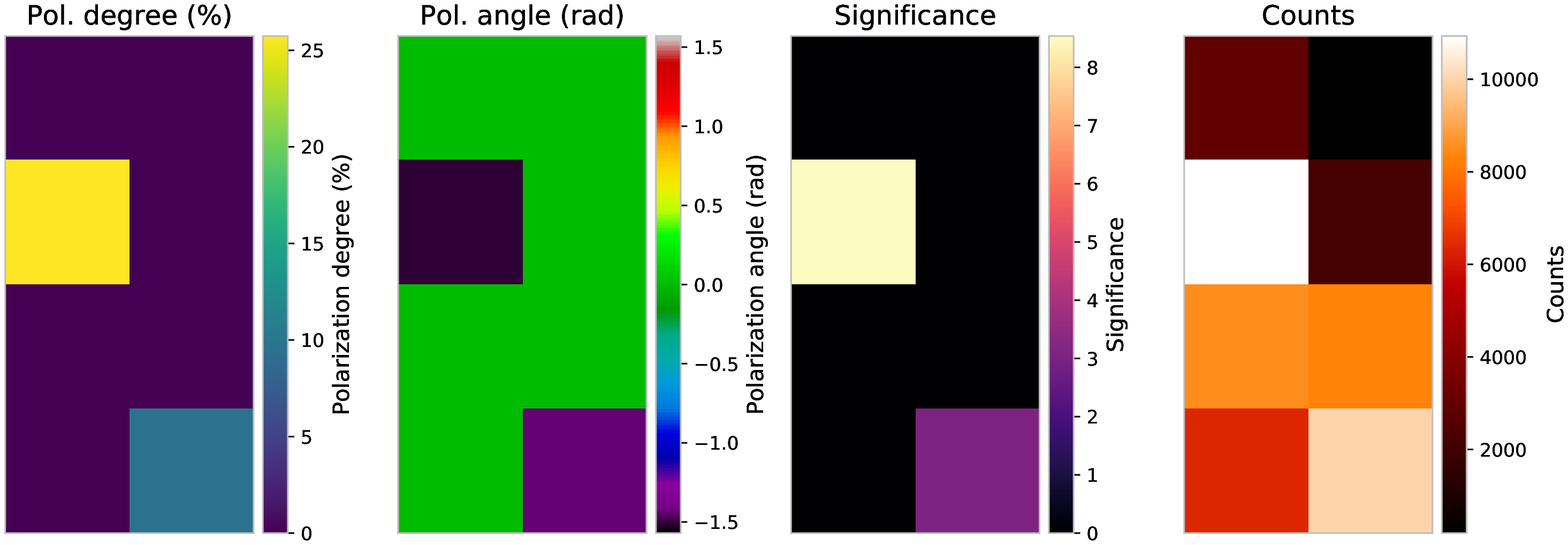}
\vfill
\includegraphics[trim={0 2.7cm 0  1.7cm},clip,width=0.5\textwidth]{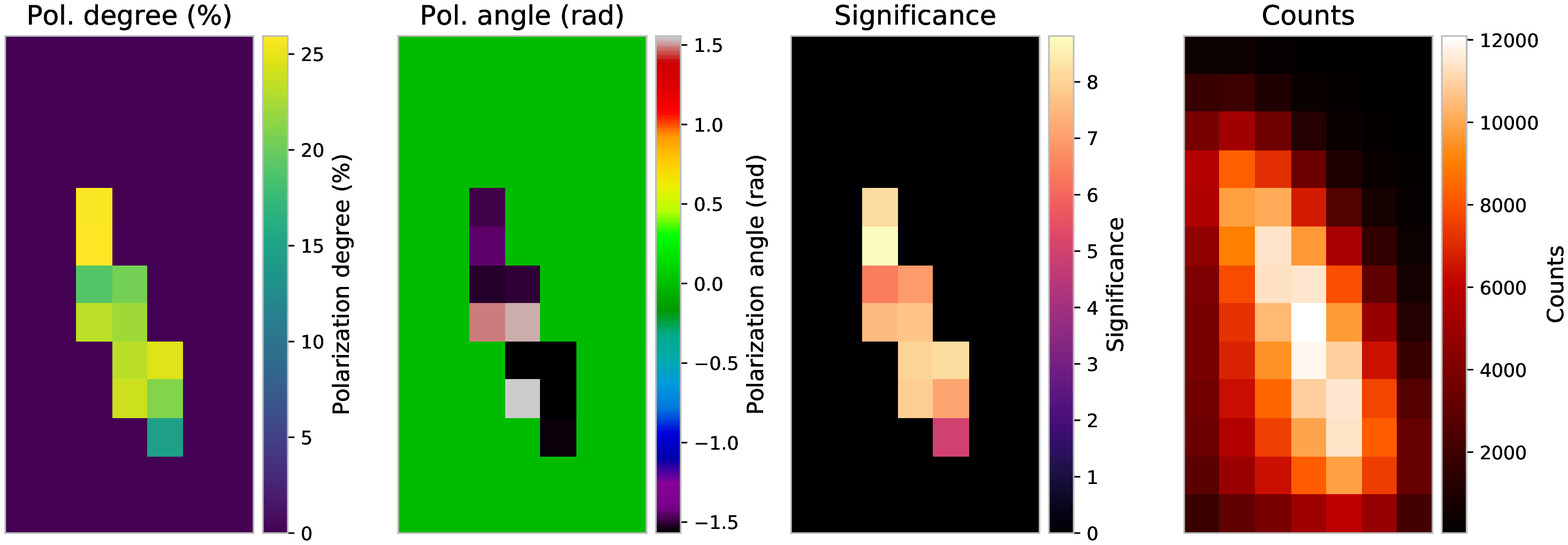}         
\caption{Top and bottom panels show {\sc ximpol} simulated maps for the case of anisotropic
turbulence produced by shock compression with $q=5$ and $\delta=5/3$.
The upper panels show maps with pixel size equal to the region from
which statistics were collected.
The lower panels were obtained for the dense grid with pixel size smaller
than the size of the circular  region that was used to collect statistics (``sliding circle'' approach).
Each row shows maps of polarization
degree, angle, significance, and total counts. The maps of polarization
degree, angle, and significance show only pixels with collected statistics
better than 10,000 counts.
\label{fig:IXPE_anisotropic-x5}}
\end{figure}

For our simulations we have used
the publicly available 
{\sc ximpol} code\footnote{https://github.com/lucabaldini/ximpol}
with the high resolution maps discussed above as sources.
For the electron flux we used an {\sc xspec} power law model 
with spectral index
$2.2$ and normalization $1.2\cdot10^{-2}$ ph/keV/cm$^2$/s at 1 keV. 
These numbers result from the
spectral fitting of the \chandra\ flux from the box region
shown in the \chandra\ image in Fig.~\ref{fig:geometry}. 
To account for the unpolarized thermal emission from the SNR, we reduced the 
model polarization degree by the ratio of the observed synchrotron flux to 
the total flux.
%
This ratio was determined by fitting joint thermal/nonthermal models
to spectra obtained from Chandra observations of Tycho. The models
provide only modest representations of the complex X-ray spectra
from the different regions, but yield reasonable estimates for the
thermal and nonthermal flux. Maps of the ratio of the synchrotron
flux to the total flux are shown in Fig. 8 for the indicated energy
bands.
%
%
To decrease the influence of the thermal plasma, we model
3--8~keV radiation for which the fraction of thermal plasma emission near the ridge is $\sim 0.2$. 

\begin{figure}          
\vskip12pt
\epsscale{1.0}
\plotone{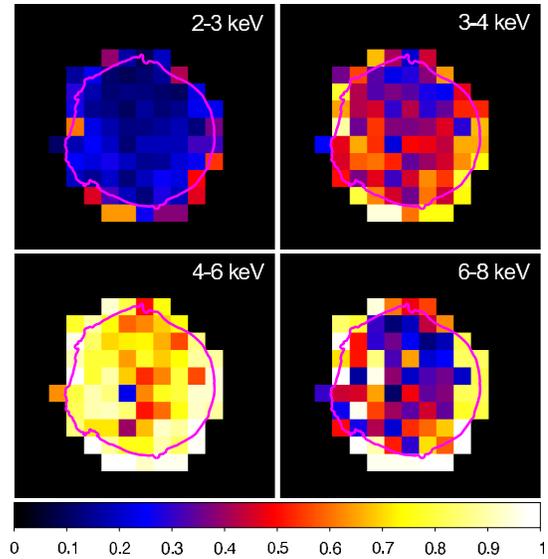}   
\caption{
Ratio maps of the synchrotron flux to total flux for Tycho's SNR.
Values from these maps were used to approximate the ratio of polarized
to unpolarized flux for the region in Fig.~\ref{fig:geometry} being
considered for simulations presented here. The outermost contour from the Chandra image of Tycho is shown for reference.
\label{fig:ratioNT}}
\end{figure}

%

Figure~\ref{fig:IXPE_anisotropic-x5} shows the results of the {\sc
ximpol} simulation for anisotropic turbulence produced by shock
compression and shown in Fig.~\ref{fig:1.2_isotr_and_anisotr} for
$q=5$ and a total exposure of 1~Ms. The upper panel shows the results
using 1 arcmin size pixels  within the image. 
Highly significant polarization is detected
in one region, with lower significance obtained in one other region.
The measured polarization angle for these regions is consistent with
that for the large-scale structures in Fig.~\ref{fig:1.2_isotr_and_anisotr},
while the polarization degree is somewhat lower, as expected from
the contributions of small disorganized regions.

To investigate spatial variations in the polarization, we have adopted
a ``sliding circle'' approach wherein we use a dense grid with 
$\approx17.5$ arcsec
size pixels and obtain
statistics from a 30-arcsec radius circle centered on the pixel. Each
pixel thus shows a numerical value that actually corresponds to counts
from a larger surrounding region. We show only pixels
with greater than 10,000 counts, sufficient to detect $\sim10\%$
polarization with 99\% c.l. (with a detector modulation factor 0.5
and without background \citep[][]{Weisskopf_2010}).

The dense grid, while showing pixel-to-pixel values that are
not statistically independent (and do not correspond to the small 
pixel size), provides an important guide to the source structure.
This is particularly true for the sharp border of the SNR, where 
an arbitrary large region used for extracting X-ray events may
fail to yield sufficient counts to detect the polarization along
the rim. 
As can be seen from comparison with
Fig.~\ref{fig:1.2_isotr_and_anisotr}, the sliding-circle maps
provide a more faithful representation of the input model for
the simulations. For a more complete analysis, a tessellation scheme
that samples the image so as to obtain sufficient 
counts to probe the polarization structure could be employed. This
is beyond the scope of our investigation here, which is aimed at
demonstrating broad sensitivity to the turbulence models described
above. 
The \ixpe\  field of view is $\sim 10^{\prime\prime}$ in diameter,
sufficiently large for single pointings to encompass many SNRs in
their entirety, including Cas A and Tycho's SNR.  
Based on the results from the upper panel, we conclude that, under the assumptions
of the associated turbulence model, a $\sim 1$ Ms \ixpe\  observation of
Tycho would yield $\sim 15$ discrete regions from the entire SNR for which
high-significance polarization detections would result.

\begin{figure}        
\vskip12pt
\epsscale{1.2}
\begin{center}{Strong cascade}\end{center}
\includegraphics[trim={0 2.7cm 0  1.7cm},clip,width=0.5\textwidth]{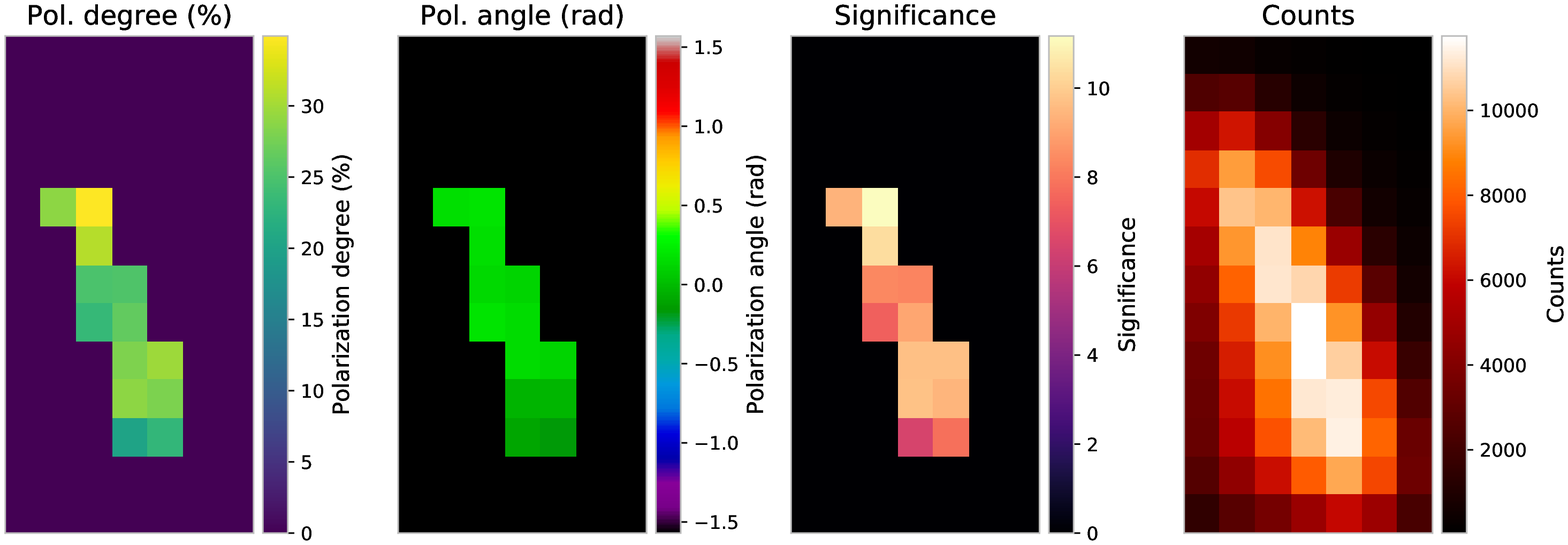}
\begin{center}{Weak cascade}\end{center}
\includegraphics[trim={0 2.7cm 0  1.7cm},clip,width=0.5\textwidth]{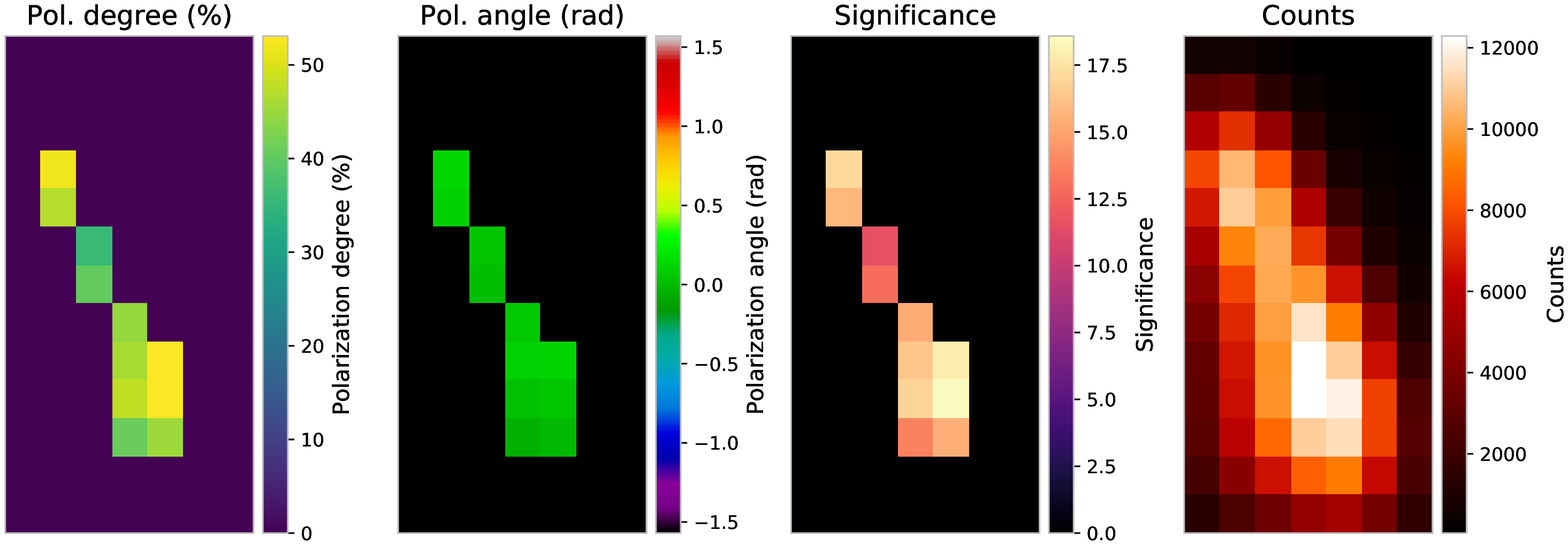}             
\caption{The upper panels show the {\sc ximpol} simulated maps for a strong cascade
case (upper row of Fig.~\ref{fig:cascade_image}) and the lower panels
show the same maps for the weak cascade case 
(lower row of Fig.~\ref{fig:cascade_image}).
Shown are the 
polarization degree, angle, significance,
and total counts (intensity) for pixels with  collected statistics greater than 10,000 counts. 
All panels are for the dense grid with pixel size smaller
than the circular region that was used to collect statistics.
\label{fig:IXPE_cascade}}
\end{figure}

Figure~\ref{fig:IXPE_cascade} shows {\sc ximpol} simulations for anisotropic cascade 
produced turbulence again assuming a 1 Ms \ixpe\ observation. 
Comparing Figs.~\ref{fig:IXPE_anisotropic-x5} and \ref{fig:IXPE_cascade} shows that 
it is possible to distinguish anisotropic
cascade turbulence from shock compressed turbulence by
the direction of polarization. 
 Indeed \citet{Strohmayer_Kallman} obtained that a
$1\sigma$ angle error in a polarization angle measurement is  
$\sim 28.5/\beta$ deg, where $\beta=a/\sigma_a$, $a$ is the polarization amplitude, 
and $\sigma_a$ is  the $1\sigma$ polarization amplitude error.
The significance value that is plotted in  
Figs.~\ref{fig:IXPE_anisotropic-x5} and \ref{fig:IXPE_cascade} 
is also equal to $a/\sigma_a$. 
Our simulations show that in some cases of the anisotropic magnetic turbulence we should 
detect polarization with significance $\sim 8$ or even greater from $\sim 15$ regions 
along the Tycho SNR shell.
This would suggest that it is possible to obtain $\sim 10$ degree, $3 \sigma$ accuracy for 
these regions that is sufficient to distinguish the cases of radial, tangential, and 
intermediate polarization directions  one from another.

In the area near the SNR blast wave the situation
is as follows: for turbulence produced by anisotropic
cascade the assumed direction of the magnetic field is along the
SNR radius and the  direction of polarization is tangential
to the SNR shock. For turbulence produced by shock compression,
the dominant magnetic field direction will be in the 
shock plane making the dominant direction of polarization along the SNR radius.
For isotropic turbulence (i.e., isotropic cascade) it is not
possible to detect significant polarization  with a 1 Ms exposure.
The {\sc ximpol} simulations are not shown for this case.

\section{Conclusions}

The stochastic properties of magnetic turbulence leave observational
traces in the small-scale structure of synchrotron X-ray intensity
and polarization maps of astrophysical objects. 
In SNR images this structure is manifested as clumps, dots, and filaments. It is sensitive
to the map energy band, to the underlying electron spectrum,  and to the level of anisotropy 
of the magnetic turbulence. Due to the characteristics of DSA in young SNRs, 
the X-ray \syn\ energy band is the most sensitive band to study this effect. 

When properly analyzed, polarization effects can provide additional information on 
the X-ray \syn\ structure beyond that obtained from  synchrotron intensity maps alone.
%
It is important that, for certain types of magnetic turbulence,
some features in the images of polarization degree and angle remain
even with the low angular resolution of 20--30 arcsec expected from 
the current generation of X-ray polarimeters.

In this article we have discussed the difference in the X-ray synchrotron
map structures that emerge in the cases of anisotropic turbulence
produced by shock compression and by cascading processes. 
%
%
%
%
%
Our results, simulating observations of $\sim 1/8$th of Tycho's SNR, suggest that a 
1 Ms \ixpe\ observation will have sufficient sensitivity to yield highly-significant 
polarization detections in multiple discrete regions around the entire SNR boundary.
If strong turbulence is short-scale and isotropic upstream of the outer blast wave, 
we will see mostly radial polarization in the downstream region after shock 
compression. However, if isotropic cascading occurs downstream, small-scale 
domains will form with random magnetic field directions and polarization that is
undetectable with a 1 Ms \ixpe\ observation. We note that, during testing  
of the radial (or tangential) polarization model, polarization angles around the 
entire periphery of the SNR can be re-phased to greatly 
increase the sensitivity of the measurements.

On the other hand, if CRs are accelerated at the forward shock efficiently enough to
 have a hard spectrum  (i.e., an index of 2 or less at the highest energies
indicative of \NL\ DSA) then we can expect 
to have strong long-wavelength fluctuations of spatial 
scales up to 10~arcsec  in Tycho. Also, the long-wavelength magnetic structures could 
be produced by the non-resonant CR mirror instability which may indicate the 
superdiffusive regime of CRs acceleration by supernova 
shocks \citep[][]{Bykov_SuperD2017}.  
Since the lifetime of large-scale fluctuations is of the order of a few  years or longer, the dominant 
magnetic field seen by a snapshot will be a random  combination of patches of quasi-regular 
strong fields. 
Within the anisotropic cascade model, when the magnetic field near the shock
is predominantly radial, this pattern will produce polarization that  is predominantly parallel to the shock front and distinct from the case of shock compressed turbulence.

Of course this picture is oversimplified since we are only considering the forward 
shock and projection effects, even for the forward shock, will tend to obscure the polarization 
direction. A clear prediction, however, comes from the expected timescale for magnetic fluctuations.
If \ixpe\ observes Tycho twice within more than a ten-year interval, significant 
changes in the polarization are expected from the anisotropic cascade, 
CR-driven turbulence models.

We have described features detectable with the new generation of
X-ray polarimeters such as \ixpe.
These detectors should give unique and 
valuable information on the properties of magnetic turbulence including its origin and 
evolution in SNRs.

\section{Acknowledgments}
The authors thank the anonymous referee for a careful reading of our paper and constructive comments. 
Most calculations were done on computers of the RAS JSCC and St-Petersburg department
of the RAS JSCC and at the “Tornado” subsystem of the St. Petersburg
Polytechnic University supercomputing center. AMB was supportet by the RSF grant 16-12-10225. 
PS acknowledges partial support from NASA contract NAS8-03060.

\newcommand{\aapDCE}{A\&A}
\bibliographystyle{aa} 
\bibliography{references5}

\end{document}